\documentclass[conf]{new-aiaa}
\usepackage[utf8]{inputenc}
\usepackage{gensymb}
\usepackage{caption}
\usepackage{subcaption}
\usepackage{graphicx}
\usepackage{amsmath}
\usepackage[version=4]{mhchem}
\usepackage{siunitx}
\usepackage{longtable,tabularx}
\usepackage{placeins}

\title{A Reduced-Order Discrete-Vortex Method for Flows with Leading-Edge Vortex Shedding}

\author{Pedro H. Gelado\footnote{PhD Candidate, p.hernandez-gelado.1@research.gla.ac.uk, Student Member AIAA}, Kiran Ramesh\footnote{Senior Lecturer, Kiran.Ramesh@glasgow.ac.uk, Senior Member AIAA.}}
\affil{Aerospace Sciences Division, School of Engineering, University of Glasgow, G12 8QQ, Glasgow, United Kingdom}

\begin{document}

\maketitle


\begin{abstract}
The formation of the leading-edge vortex (LEV) is a key feature of unsteady flows past aerodynamic surfaces, but is expensive to model in high fidelity computations. Low-order methods based on discrete vortex elements are able to capture the physical behavior of these flows, in particular when enhanced with a criterion that models the ability of the leading edge to sustain suction. These models are significantly faster than high order methods, but their expense still grows as vortex elements are continuously shed and convected into the wake, in effect an $\mathcal{O}(n^2)$ problem. This work proposes accelerating the leading-edge suction parameter discrete vortex method (LDVM) by limiting the number of vortex elements in the LEV coherent structure to N, hence giving the name to the method N-LEV LDVM. The N-LEV LDVM method correctly approximates the flows in comparison with the original LDVM model and computational fluid dynamics (CFD) simulations until the point of LEV detachment, which N-LEV LDVM is unable to model. We propose reintroducing this behavior via two physical detachment criteria studied in LEV literature, a threshold of maximum circulation in the LEV and trailing edge flow reversal. We demonstrate the ability of the N-LEV LDVM method to accurately predict the instant in time this detachment occurs for both mechanisms in comparison with experimental results, laying the ground for their incorporation into the method.
\end{abstract}

\section*{Nomenclature}

{\renewcommand\arraystretch{1.0}
\noindent\begin{longtable*}{@{}l @{\quad=\quad} l@{}}
$A_0$  & leading-edge Fourier coefficient \\
$A_n$  & n-th Fourier coefficient \\
$c$  & chord length \\
$C_d$  & drag coefficient \\
$C_l$  & lift coefficient \\
$C_m$  & lift coefficient \\
$\dot{h}$  & plunge rate \\
$t$  &   time \\
$U$  & free stream velocity \\
$W$  & downwash \\
$\alpha$  & pitch angle \\
$\dot{\alpha}$  & pitch rate \\
$\alpha_{eff}$  & effective pitch angle \\
$\gamma$  & vorticity distribution \\
$\theta$  & transformation variable \\
$\phi$  & velocity potential \\
\end{longtable*}}

\section{Introduction}
 Airfoils undergoing unsteady kinematics cause the formation of a Leading Edge Vortex (LEV) that is the source of strong forces and moments that underpin biological flight\cite{ellington1984aerodynamics}\cite{ellington1999novel}. Experimental results have shown rich features occurring in these flows, with eddies, apparent-mass effects and boundary layer separation causing non-conventional routes to generate lift\cite{eldredge2019leading}. This has caused its research interest to span to a plethora of fields, from improving efficiency of energy capturing devices\cite{pisetta2022morphing}, through helicopter rotor dynamics\cite{mulleners2012onset}, to the design of bio-mimicking flapping swimmers and fliers\cite{karasek2018tailless}. Understanding, modeling and utilizing these flows is hence a central task to tackle sustainable development goals, and yet still present significant challenges. 
 
 Unsteady flow modeling methods can be roughly categorized into classical solutions, numerical methods and data-driven models. Classical closed-form approaches developed by Theodorsen \cite{theodorsen1935general}, Wagner \cite{wagner1925dynamischer}, Garrick \cite{garrick1937propulsion} or von Karman \& Sears \cite{sears1939growth} offer analytical solutions at the expense of simplifications that constrain solutions, as underlying assumptions restrict applicability in cases with detached flow or large, rapid effective angle of attack variations. High-fidelity numerical methods expand the reliability of models in more exotic kinematics, but having to solve for a fourth dimension, time, and allowing for more complex geometries, causes the computational cost of modeling unsteady flows to highly exceed that of its steady counterparts \cite{ferziger2002computational}.

This has led to the birth of a variety of approaches to create low-order models and build towards real-time performance. This goal is essential in control of flapping autonomous vehicles, sensor-in-the-loop state estimation, and live prediction of the behavior of physical assets like wind farms or wave energy collectors. 

Discrete-vortex methods show promise by using the same vortex interactions that increase the complexity of these flows as the tool to model them. The flow field is hence recast as a set of discrete parcels of vorticity that are generated independently from the surface of the airfoil, and which  behave according to classical vortex dynamics \cite{darakananda2019versatile}.

Further dimensional reduction can be achieved by limiting the number of vortex elements in the field. This significantly reduces computational cost, as shown in methods developed by Sarpkaya \cite{sarpkaya1975inviscid}, Chorin \cite{chorin1990hairpin} and more specifically in the merging of vortex elements minimizing accuracy loss by Rossi \cite{rossi1997merging}. 

Experimental work on LEV formation and dynamic stall point out three qualitatively distinct phases in its evolution: initiation, growth and shedding of the LEV \cite{mulleners2012onset}. Whilst original discrete-vortex methods either continuously shed new elements from the leading edge or did not at all\cite{katz_plotkin_2001}\cite{cottet2000vortex}, Ramesh et al.\cite{ramesh2014discrete} proposed a leading-edge suction parameter (LESP) to trigger the shedding of vortex elements from the leading-edge. This modeled the degree of suction the rounded leading-edge of an airfoil is able to sustain, and related it to the $A_0$ term of unsteady thin airfoil theory\cite{katz_plotkin_2001}\cite{ramesh2011augmentation}\cite{eldredge2019leading}. This LESP criterion was then integrated into unsteady thin airfoil theory discrete vortex method as proposed by Katz and Plotkin\cite{katz_plotkin_2001}. Moreover, recent developments in LESP theory point that the criterion may in fact not be a constant but a function of the velocity at the shear layer as evidenced by experimental results \cite{martinez2022modulation}. 

As we will show in this work, in Ramesh et al.'s original LDVM method\cite{ramesh2014discrete} initiation and growth of the LEV is not affected by limiting the number of vortex elements in the LEV coherent structure, as vorticity is still entrained as vortex elements are merged, causing the LEV to initiate via the LESP and grow in size. However, the rich interactions between the vortex elements shed from the leading edge (LE) and trailing edge are lost, hence the vortex sheet formed at the LE is unable to roll-up and the LEV detachment mechanism is lost, making limiting the number of vortex elements in the method nonphysical. 

How can we capture the richness of the interactions between "big whirls" and their "lesser whirls" \cite{richardson2007weather} to reintroduce this mechanism?

Darakananda and Eldredge \cite{darakananda2019versatile} proposed a useful taxonomy for reduced discrete vortex methods using two parameters $B_f$, a constant that represents the tolerance for error in the model, and $T_{min}$ the minimum time interval between the release of new vortex elements.  If we let $T_{min} = 0$, the model is pure discrete vortex method, as in every time step a vortex will be shed. However, if we let $T_{min} = \infty$ and $B_f=0$, no new vortex elements will be generated at all during the model's simulation time, as the time interval between their generation is equal to infinity, hence giving us a vortex sheet model. If both $T_{min} = \infty$ and $B_f=\infty$, the model closely resembles the impulse matching model proposed by Wang and Eldredge \cite{wang2013low}. Values in between these create a variety of different hybrid models under which those proposed by Pullin\cite{pullin1978large}, Jones\cite{jones2003separated} and Li and Wu \cite{li2016vortex} could fall. 


Moreover Darakananda and Eldredge \cite{darakananda2019versatile} proposed a hybrid approach with vortex sheets for the LEV and TEV shear layers, and active point vortices for LEV and TEV vortex cores and inactive (fixed strength) point vortices in the wake. To create this hybrid model an algorithm parameterized by $B_f$ and $T_{min}$ was put forward, with an additional velocity correction parameter. This is computed for the vortex elements that make the vortex sheets at the leading and trailing edges of the airfoil. 

The necessity for this velocity correction stems from the fact that when merging vortex elements together, a nonphysical force is created on the airfoil surface by the transfer of circulation between these elements. This extends the original impulse matching technique developed by Wang and Eldredge\cite{wang2013low} to models with both variable strength vortex sheets and fixed or variable point vortex elements. 

Additionally, when elements are merged from the free end of the vortex sheet, representing the shear layer of the airfoil, instabilities can be caused by diverging distances between elements as circulation is absorbed by the point vortex core. To tackle this Darakananda and Eldredge \cite{darakananda2019versatile} remap the position of the elements in the vortex into a uniform distribution along its length and use Fourier filtering to remove components with wavelengths below a cutoff threshold. To avoid this nonphysically curtailing the wrapping up of the vortex sheet in small scales below the threshold as identified by Krasny\cite{krasny1986study}, Fourier filtering is only applied in cases were both a vortex sheet and active vortex element are present.


\begin{figure}[hbt!]
\centering
\includegraphics[width=.5\textwidth]{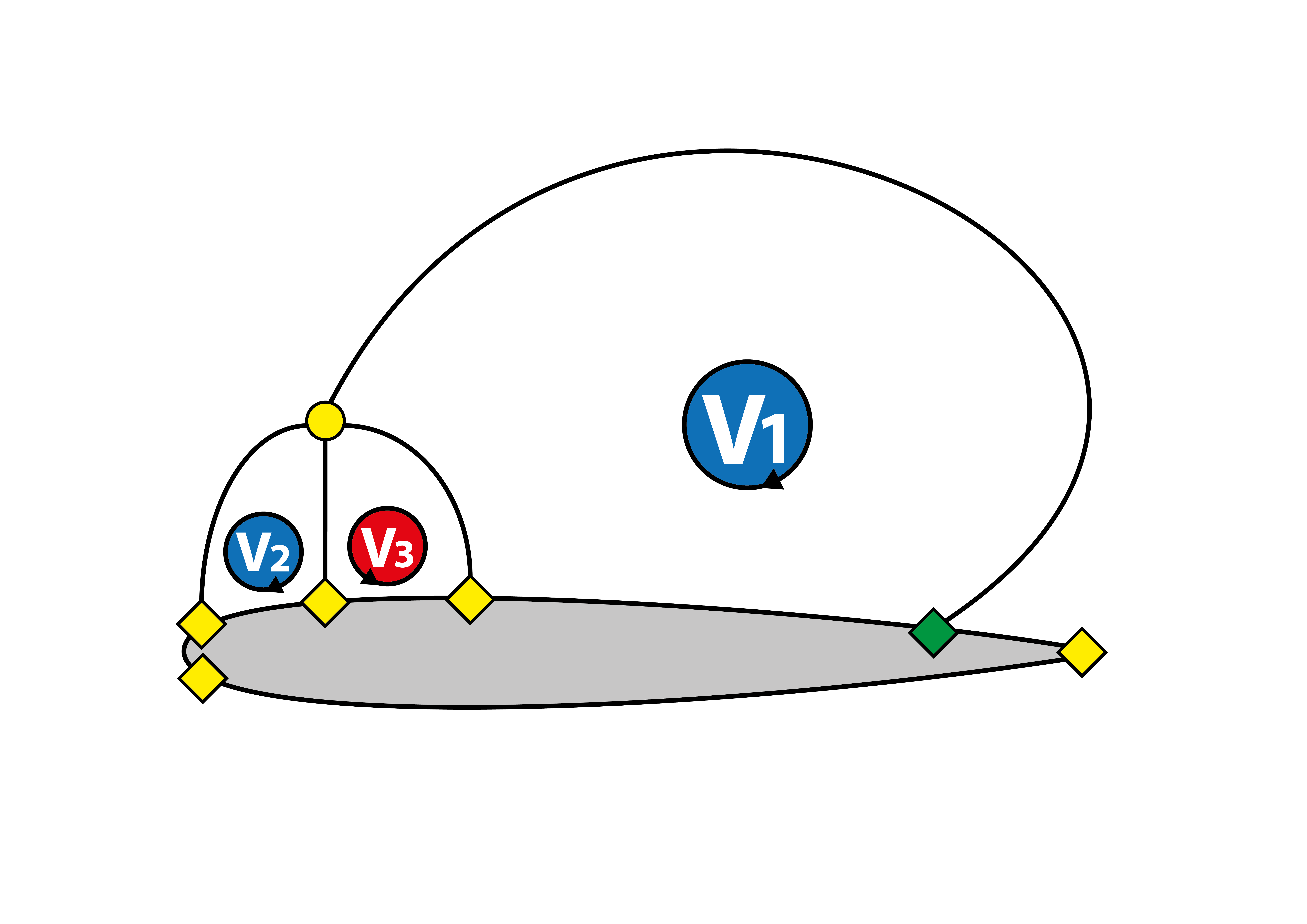}
  \caption{Main topological flow features of the LEV ($V_1$) and its secondary structures ($V_2$ and $V_3$), adapted from Kissing et al.\cite{kissing2020insights}, half-saddles/stagnation points are shown as diamonds and the main LCS saddle is shown as a yellow circle. The rear stagnation point/half-saddle is shown as the green diamond that terminates the LEV.}
    \label{fig:flowtopology}
\end{figure}

However, recent work by Kissing et al. \cite{kissing2020insights}\cite{kissing2020leading} and Widmann \& Tropea \cite{widmann2015parameters} on the topology of the leading edge vortex has uncovered several mechanisms by which it detaches from the airfoil surface. This creates an opportunity to enhance element constrained discrete vortex methods. The n-vortex approach presented in this paper will build on the LDVM method developed by Ramesh et al. \cite{ramesh2014discrete}, by constraining the number of vortex elements to increase computational speed. Moreover, a route to LEV detachment is incorporated via the mechanisms uncovered by Kissing et al. and Tropea et al. \cite{kissing2020insights}\cite{kissing2020leading}, built from earlier work with Widmann et al.  \cite{widmann2015parameters}. 


Of special interest will be kinematics that characterize unsteady flows in the low Reynolds number regime applicable to Unamnned Aerial Vehicles (UAVs), energy capturing devices and insect flight. The novel approach presented in this paper is called N-LEV LDVM, as the number of vortices shed from the leading edge of the airfoil present in the LEV structure is limited to a maximum value of N.

\section{LESP-modulated discrete-vortex method (LDVM)}

The LDVM method was formulated by Ramesh et al. \cite{ramesh2013unsteady}\cite{ramesh2014discrete}\cite{ramesh2011augmentation} as an extension of unsteady thin airfoil theory developed by Katz and Plotkin \cite{katz_plotkin_2001}, to include leading edge vortex shedding via a leading edge suction parameter (LESP) and extend the method beyond small angle assumptions. Below the method is briefly summarized from Ramesh et al.'s seminal work on the method, the interested reader is encouraged to consult the original paper for further details \cite{ramesh2014discrete}.

The vorticity distribution $\gamma(x)$ along the chord-wise coordinate is taken as a Fourier series:

\begin{equation}
    \gamma(\theta,t) = 2U\left[A_0(t)\frac{1+cos\theta}{sin\theta} + \sum_{n=1}^{\infty} A_n(t)sin(n\theta) \right]
\end{equation}

\noindent with $\theta$: 

\begin{equation}
    x = \frac{c}{2}(1-cos\theta)
\end{equation}

\begin{figure}[h!]
    \centering
    \includegraphics[clip= true, trim={4cm 12cm 4cm 4cm},  width=0.5\textwidth]{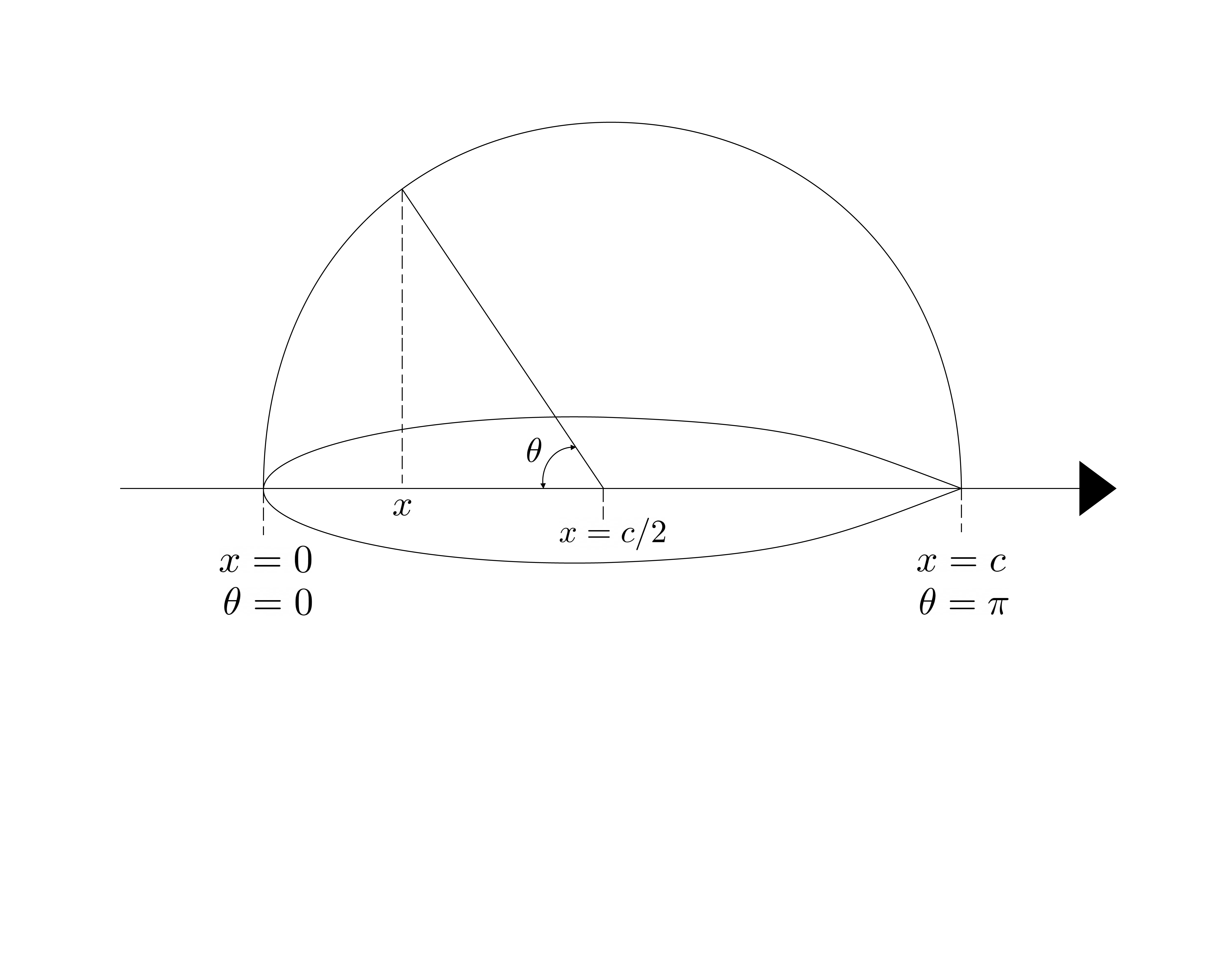}
    \caption{The airfoil transformation}
    \label{fig:airfoilz}
\end{figure}

\noindent where c is the airfoil chord, and U the velocity component in the $-X$ direction. Fourier coefficients are calculated as functions of the local downwash on the airfoil, $W(x,t)$ with the no-penetration boundary condition making the velocity tangential to the airfoil surface. 

\begin{equation}
    A_0(t) = -\frac{1}{\pi}\int_{0}^{\pi} \frac{W(x,t)}{U} \,d\theta\
\end{equation}

\begin{equation}
    A_n(t) = \frac{2}{\pi}\int_{0}^{\pi} \frac{W(x,t)}{U} cosn\theta \,d\theta\
\end{equation}

The instantaneous local downwash, $W(x,t)$, is computed form the components of the airfoil's kinematics and the velocities induced by the vortex elements in the field. Kelvin's circulation theorem is enforced by calculating the strengths of the vortex elements such that:

\begin{equation}
    \Gamma_b(t) + \sum_{m=1}^{N_{tev}} \Gamma_{tev_{m}} + \sum_{n=1}^{N_{lev}} \Gamma_{lev_{m}} = 0
\end{equation}

\noindent where the bound circulation $\Gamma_b$ is: 

\begin{equation}
    \Gamma_b = Uc\pi\left[A_0(t) +\frac{A_1(t)}{2}\right]
\end{equation}

A Newton-Raphson iteration is carried out every time step to compute these conditions. 

The LESP is defined as a non-dimensional measure of the suction at the leading edge, and is derived to be the first Fourier coefficient: 

\begin{equation}
    LESP(t) = A_0(t)
\end{equation}

A critical value of LESP determines the initiation of LEV formation. If the motion of the airfoil or vortex element interactions causes the LESP to exceed the critical value, the LE will shed vortex elements to counterbalance this effect. The strength of the shed LEVs is calculated such that: 

\begin{equation}
    LESP(t) = LESP_{crit}
\end{equation}

\noindent during the vortex shedding. 

Forces, moments and pressures can hence be derived from velocity potentials, $\phi$, which include three components stemming from  bound vorticity, LEV elements and TEV elements. 

\begin{equation}
    \phi = \phi_b + \phi_{lev} + \phi_{tev}
\end{equation}

All vortex elements in the flow field are then convected, according to the effect on each of them of the induced velocities of all the other elements in the field. The method then moves to the next time step and the process restarts. A detailed derivation of the method can be found in Ramesh et al.\cite{ramesh2014discrete}.  

\section{N-LEV LDVM}

In N-LEV LDVM, Ramesh et al.'s original LDVM is first sped up by limiting the number of vortices shed from the leading edge (LE) to a preset constant, N, which gives name to the method. When a new vortex element is generated from the leading edge in the LDVM method, the two oldest LE vortex elements in the field are merged. 

This reduces the expense of the problem by limiting its order, whilst simultaneously preserving the shear layer unlike models where a single point vortex is used. However, rich interactions occurring between vortex elements in the near-field of the airfoil that cause the detachment of the leading edge vortex is lost. This is because the merged vortex will keep ingesting vorticity generated at the leading edge indefinitely, not allowing the formation of new LEVs while old ones are shed into the wake. 

To introduce back this physical behavior lost in constraining the number of LEV vortex elements, a recently discovered leading edge vortex detachment mechanism is studied. Kissing et al. \cite{kissing2020insights}, and Widmann \& Tropea \cite{widmann2015parameters} have shown that trailing edge recirculation triggers LEV detachment. 

Hence if the rear stagnation point of the LEV, shown as a green diamond in Fig. \ref{fig:flowtopology}, reaches the trailing edge, the detachment condition is met. At this point, the merged vortex should not be allowed to keep ingesting vorticity from the LE. Instead a new LEV, with N-vortex elements should begin to be created. The original merged vortex would be hence shed into the wake as would occur physically.

The scheme proposed above would differ from others based on fixed order vortex sheet, vortex blob and their hybrids by adding a mechanism for LEV detachment \cite{eldredge2019leading}, which extends the range of validity of the method to a wider range of physical flows.

In this paper, results from the N-LEV method without the detachment condition are presented. It it shown that the method performs well until the instant of LEV detachment from the airfoil surface. Methods for identifying LEV detachment from the N-LEV method are then investigated.

%
%
%


\section{Results \& Discussion}
The Results section will be structured with initial results of the N-LEV method as currently implemented showing its current limitations, with comparisons made to the original LDVM method and CFD cases by Ramesh et al.\cite{ramesh2014discrete}.Then follows an investigation into two leading edge vortex detachment criteria found in literature \cite{widmann2015parameters}\cite{rival2014characteristic}\cite{mulleners2012onset}\cite{kissing2020insights} that are proposed to act as triggers of LEV detachment in the N-LEV method: flow reversal at the trailing edge and maximum accumulation of circulation in the LEV. Finally a discussion of challenges and the plan to combine the detachment criteria with the current N-LEV formulation will be discussed. 

The same kinematic cases used to validate the original LDVM method will be used to validate this reduced order augmented N-LEV method, as seen in Table \ref{table:kinematics}, allowing for comparison with computational and experimental data from Ramesh et al. \cite{ramesh2014discrete}. In this paper Eldredge’s canonical formulation for smoothed ramp functions will be used, the interested reader can find these in \cite{wang2013low} \cite{ol2009high}. For all simulations presented of the LDVM and N-LEV models, the time step was set to $\Delta t=0.015c/U$. The three kinematics chosen to be presented below all exhibit continuous LEV shedding. 


\begin{table}
\begin{center}
\begin{tabular}{ c c c c c c }
 Case & $\alpha_{max}$ (deg.) & $K_{\alpha}$ & Pivot $(x_{p}/c)$ & $(h/c)_{max}$ & $K_h$\\ 
 \hline
 A & 45 & 0.2 & 0 & - & - \\  
 B & 25 & 0.2 & 0 & - & - \\
 C & 90 & 0.4 & 1.0 & - & - \\
\end{tabular}
\end{center}
\caption{Motion kinematics to validate the N-LEV LDVM method against unconstrained LDVM and high fidelity simulation data.}
\label{table:kinematics}
\end{table}

\FloatBarrier

\subsection{Demonstration of N-LEV LDVM for Case A} \label{sec:pitchramp45}

\begin{figure}[h!]
\centering
\includegraphics[width=.49\textwidth]{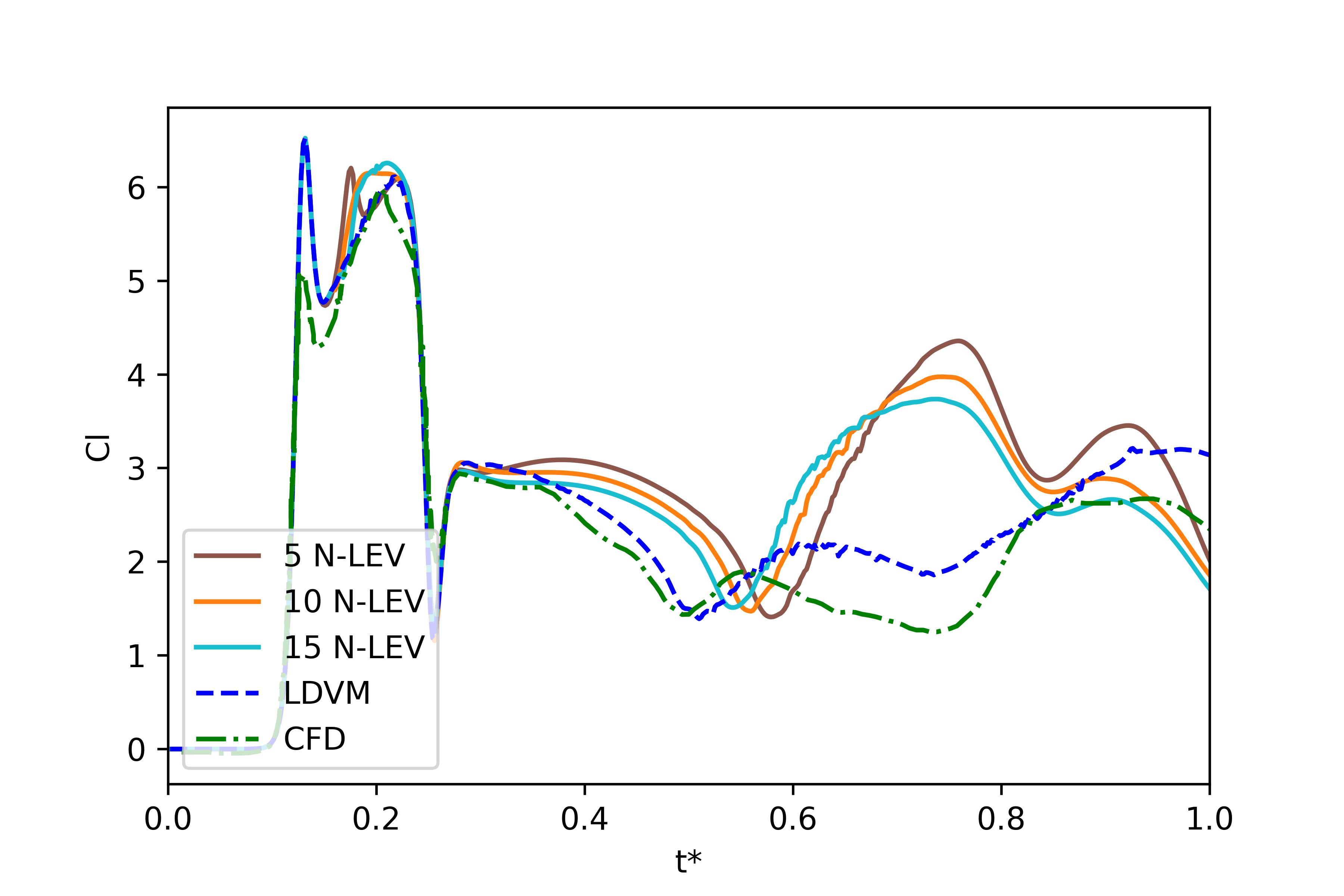}
\includegraphics[width=.49\textwidth]{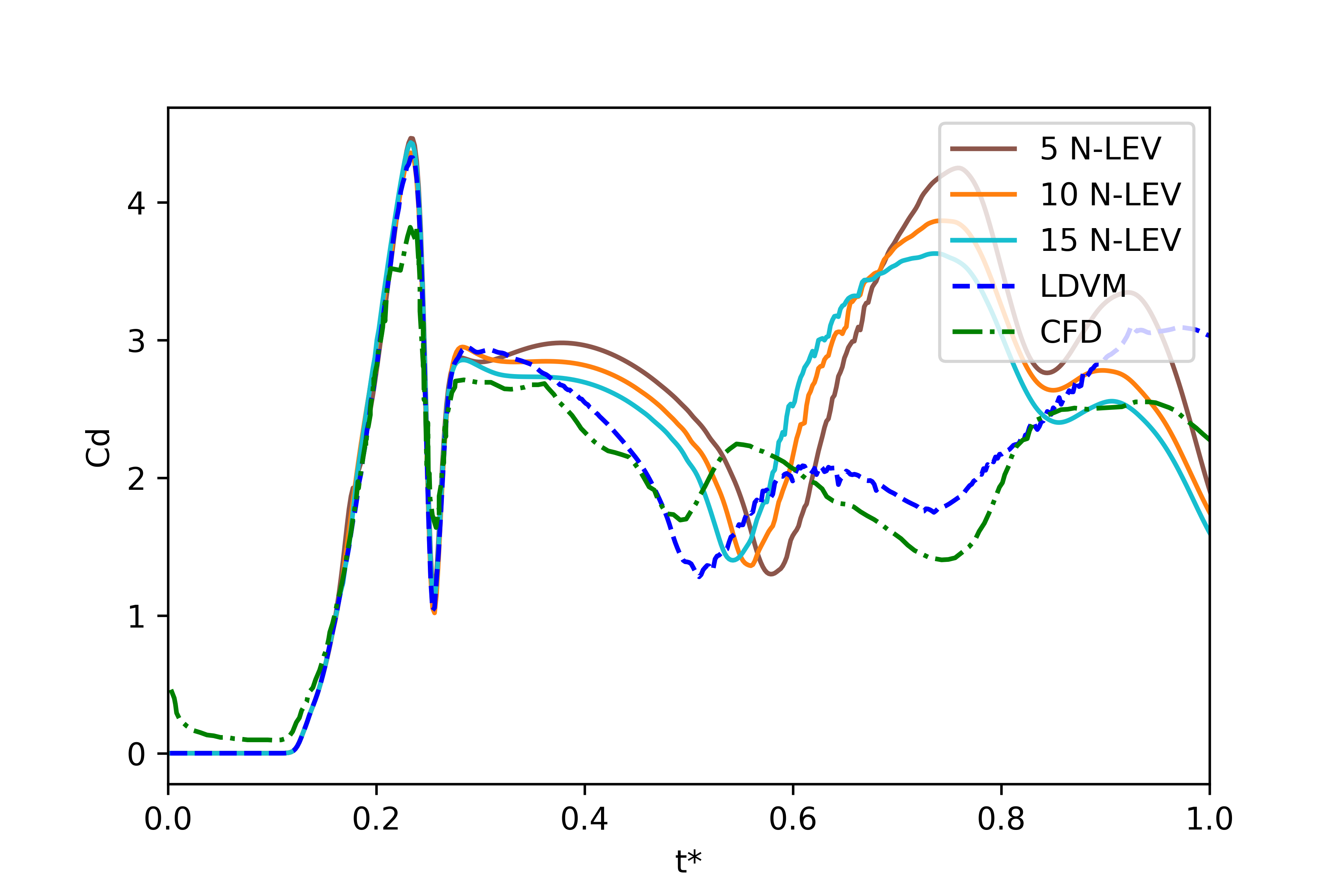}
\includegraphics[width=.49\textwidth]{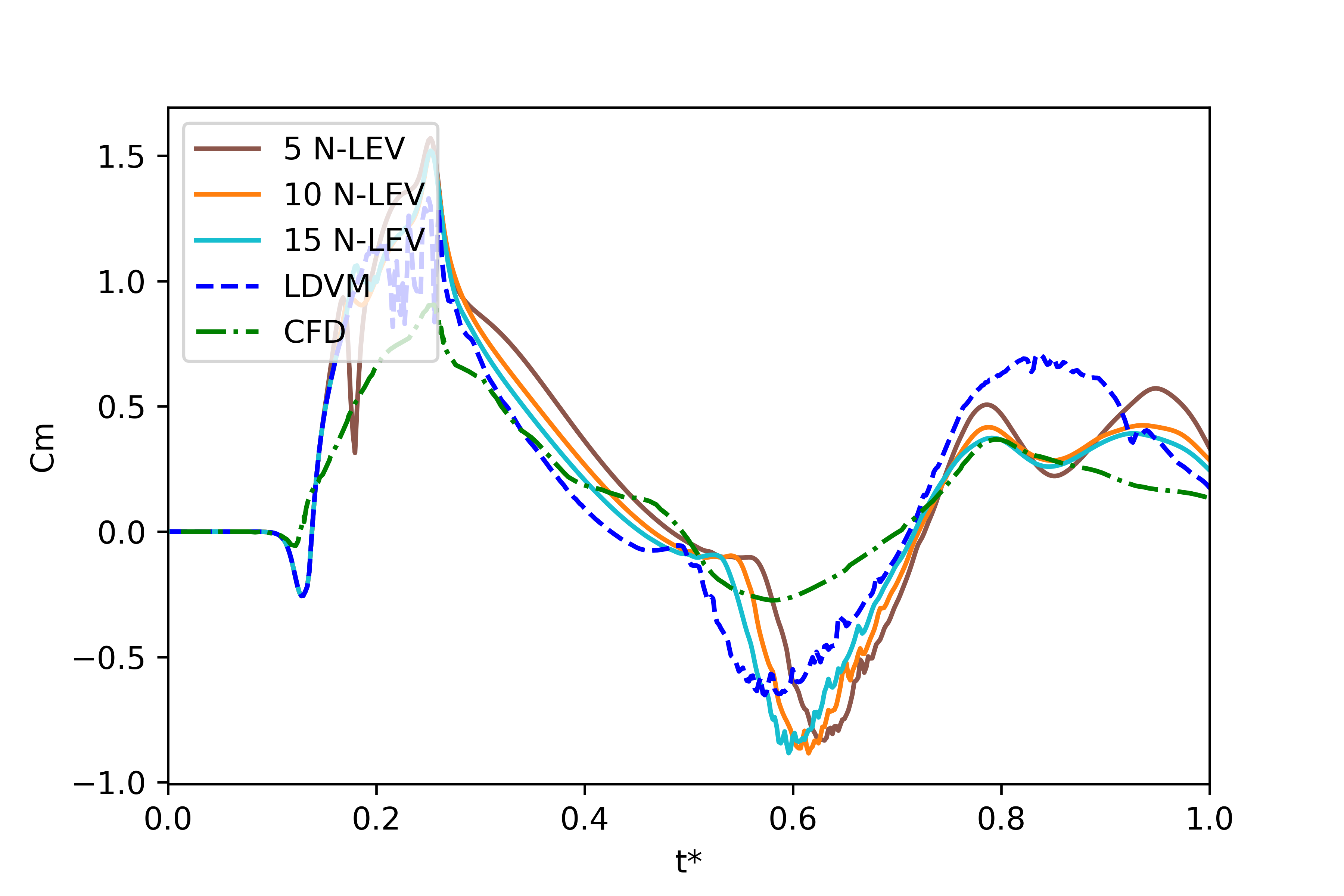}
\includegraphics[width=.49\textwidth]{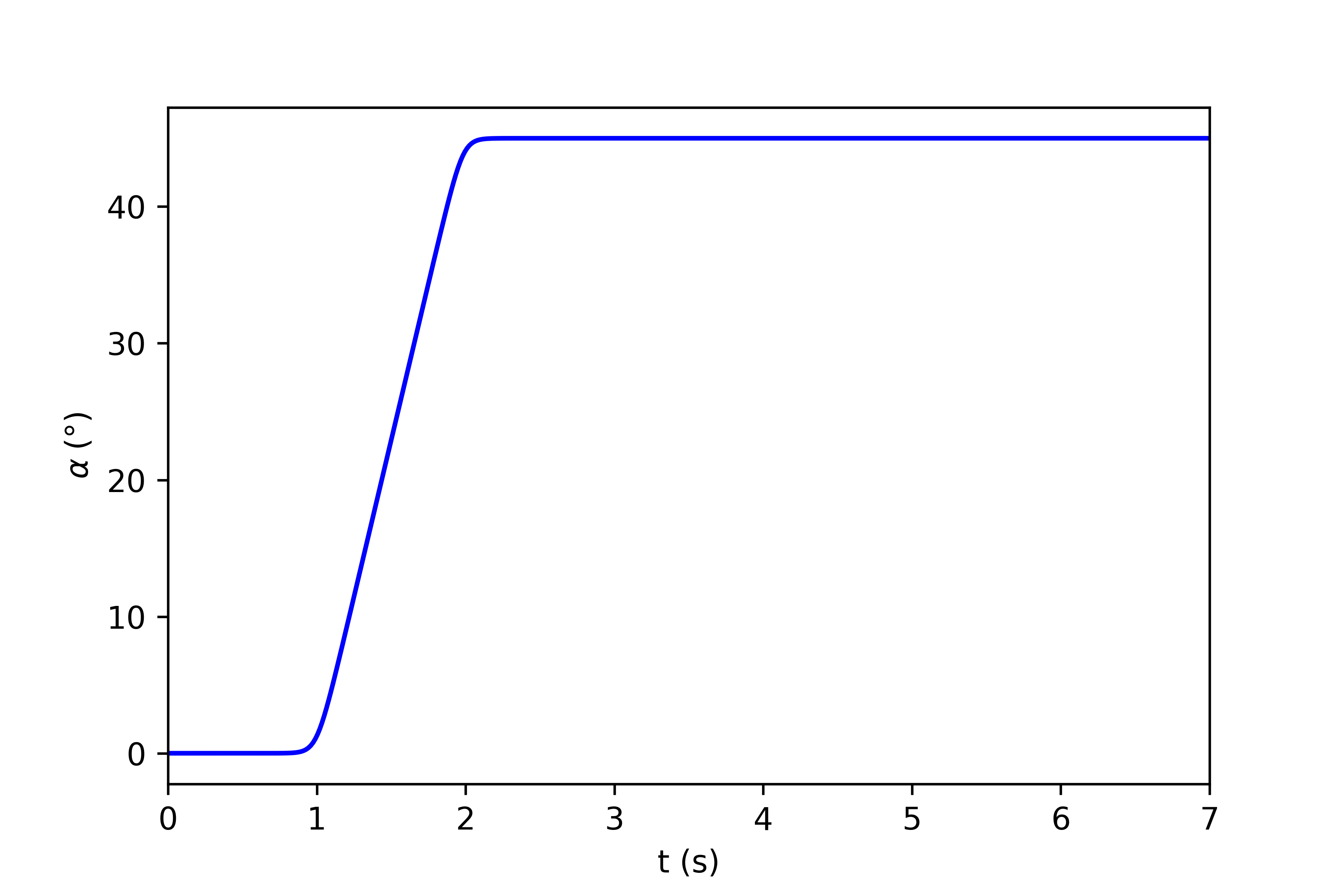}
  \caption{Case A: Comparison of $C_l$, $C_d$, $C_m$ and pitch angle from N-LEV LDVM (with 5, 10, 15 elements), classical LDVM and CFD.}
    \label{fig:resultskinem1}
\end{figure}

\begin{figure}[h!]
\centering
\includegraphics[width=.32\textwidth]{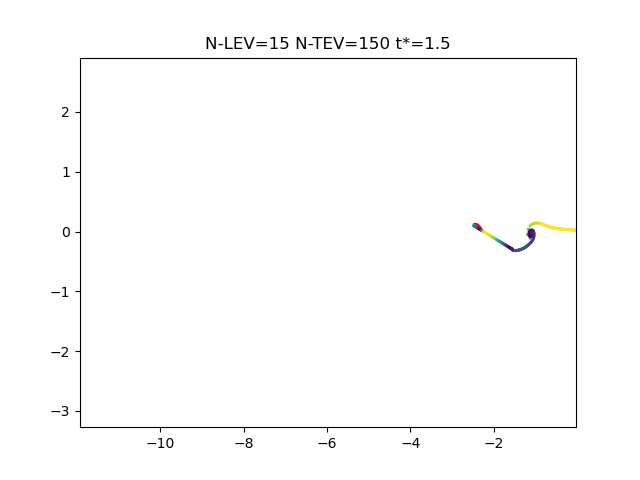}
\includegraphics[width=.32\textwidth]{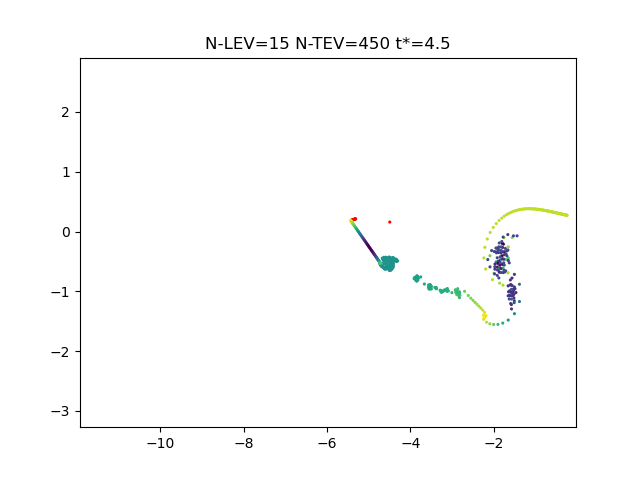}
\includegraphics[width=.32\textwidth]{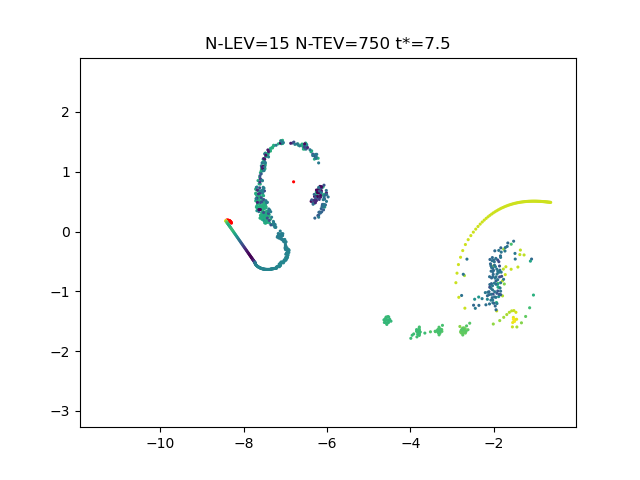}
\includegraphics[width=.32\textwidth]{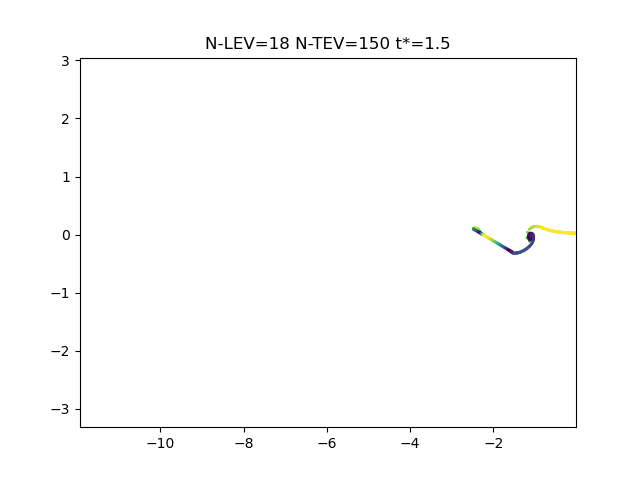}
\includegraphics[width=.32\textwidth]{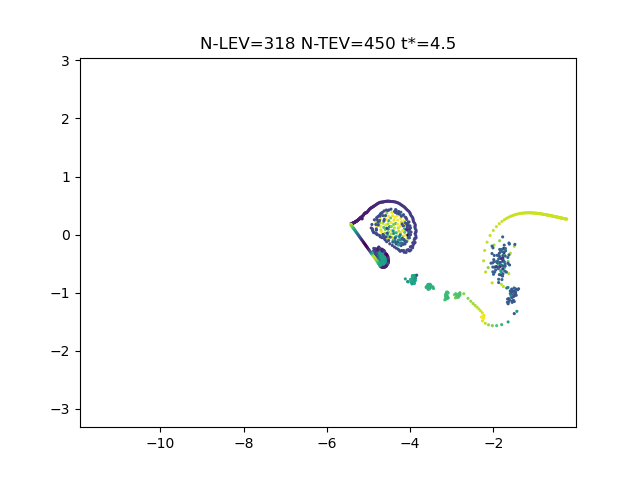}
\includegraphics[width=.32\textwidth]{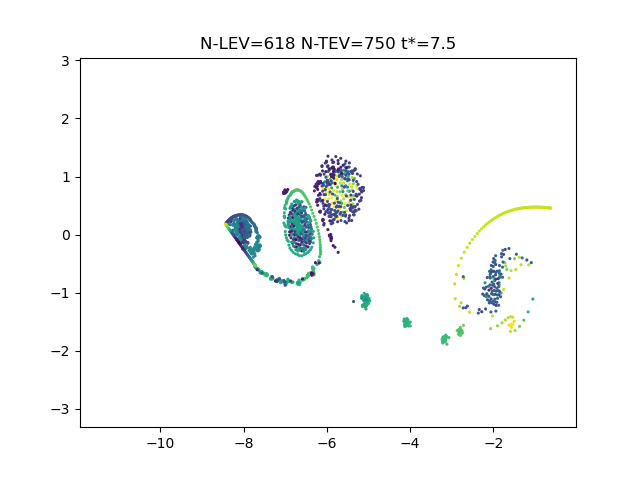}

  \caption{Case A: Vortex plots for N-LEV LDVM (with 15 elements) vs classical LDVM. The 15 LEV vortex elements are colored in red.}
    \label{fig:vortexplotskinem1}
\end{figure}

In this first example we will attempt to limit the number of vortices shed in the LDVM method from the leading edge, without constraining those from the trailing edge. Once the maximum number of vortices, N, are shed from the leading edge, "oldest" vortex in the field, i.e. the first one shed in the simulation, is merged with its second-oldest neighbor. This is repeated every time step just after a new vortex element is added to the field. 

The choice of a threshold of vortex elements for the N-LEV method can be an arbitrary integer, although the authors have found that as $t*$ increases larger numbers of vortex elements cause instabilities in the solution due to non-physical behavior in vortex merging. 

The kinematics are of an Eldredge pitch ramp-up to 45\degree, with no plunge.  We can see near exact agreement between the LDVM and the n-vortex LEV method until approximately $t^* = 3.2$, whereupon performance degrades as the large scale feature evolution occurring from vortex-element interactions in the LDVM method are not captured by the 15-element, 10-element and 5-element N-LEV approaches taken here.


Fig. \ref{fig:vortexplotskinem1} shows the vortex element plots for the LDVM and 15-element N-LEV method, and we can see the close resemblance between the N-LEV LDVM and the original LDVM method in the ramp up, at $t^*=1.5$. However, as the trailing edge vortex begins to roll-up, the single core LEV is unable to capture the rich interactions that cause its detachment. 

Moreover, at $t^*=7.5$ the trailing edge vortex sheet in the N-LEV LDVM is wrapped around this merged vortex, straying away from the rich dynamic interactions between four vortex element cumuli seen in the LDVM method. Additionally at  $t^*=7.5$ we can see a new LEV is being formed in the leading edge in the LDVM method, behavior which is absent from N-LEV LDVM. All of these differences in behavior stem from the excessive accumulation of vorticity at a non-physical location by the final element that should resemble the LEV core.

\FloatBarrier

\subsection{Demonstration of N-LEV LDVM for Case B}

As a second case study we will look at the performance of simple N-LEV limiting in the same manner as in Section \ref{sec:pitchramp45}, for a pitch-ramp-return kinematic, with no plunge. The maximum amplitude is $\alpha_{max} = 25\degree$, and the pivot location the leading edge. The LESP criterion for this kinematic was taken from Ramesh et al.\cite{ramesh2014discrete} as $LESP=0.18$, and the pitch-ramp was modeled using an Eldredge ramp-up with smoothing parameter a=11, comparison to CFD results from Ramesh et al.\cite{ramesh2014discrete} are shown. 

\begin{figure}[h!]
\centering
\includegraphics[width=.49\textwidth]{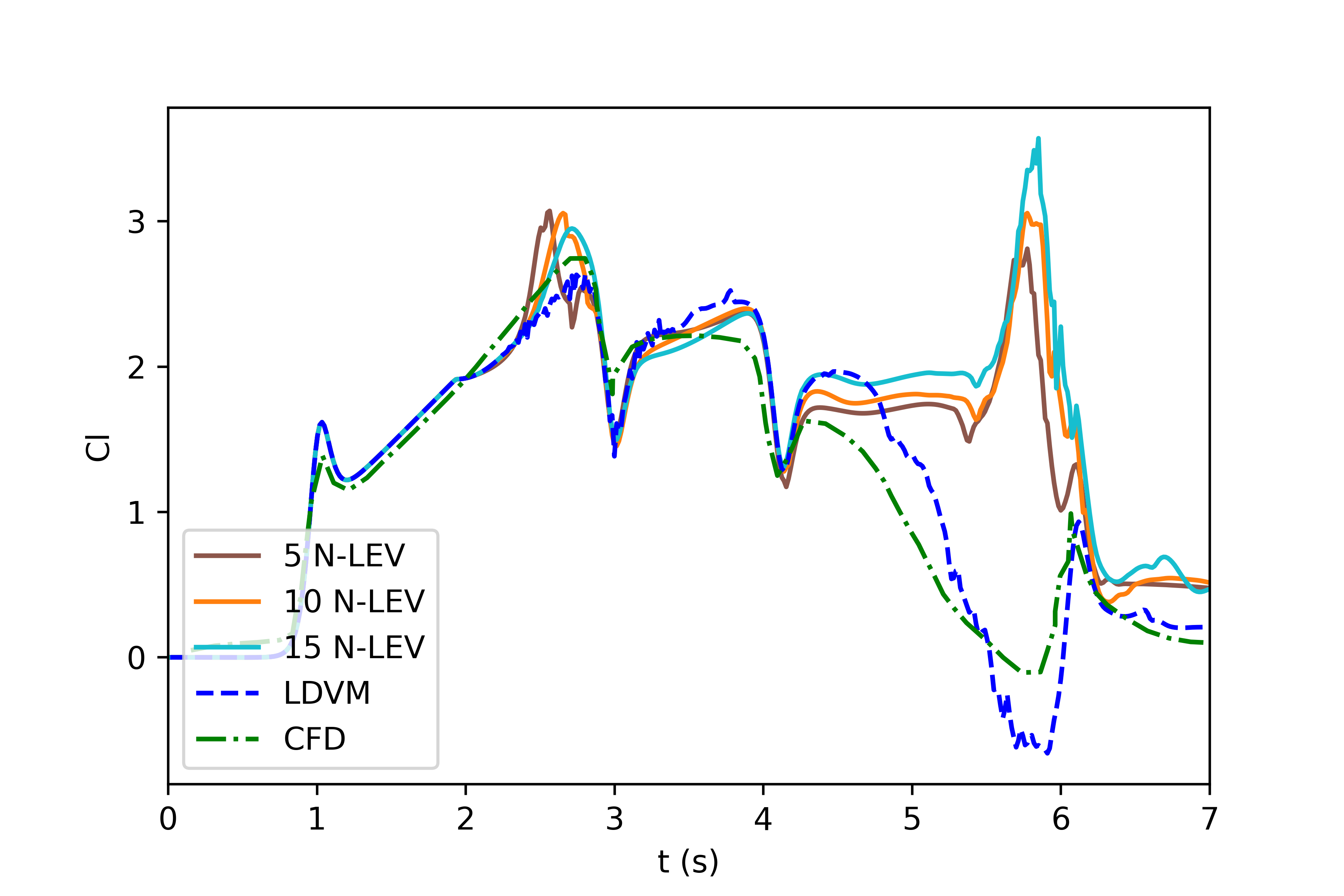}
\includegraphics[width=.49\textwidth]{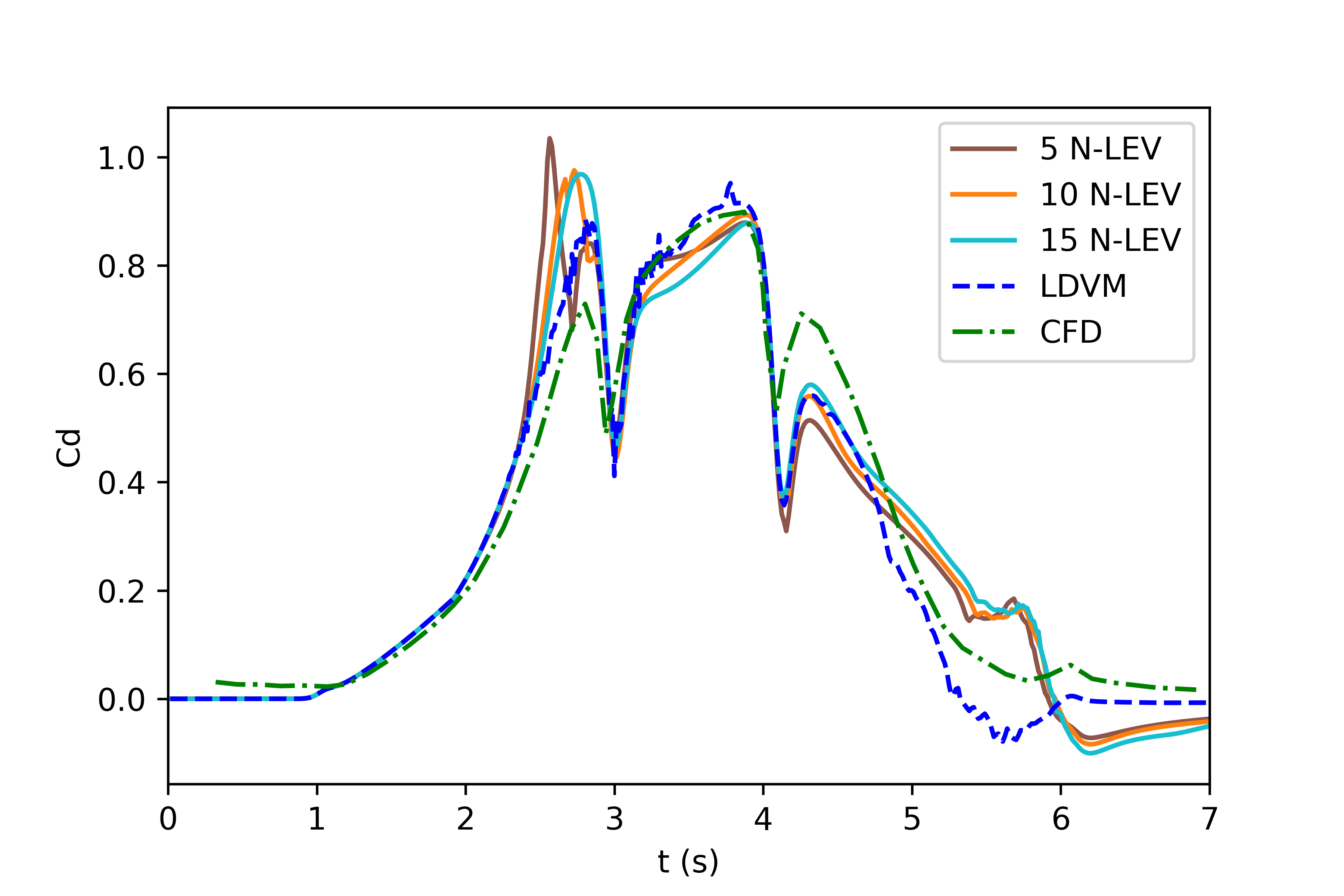}
\includegraphics[width=.49\textwidth]{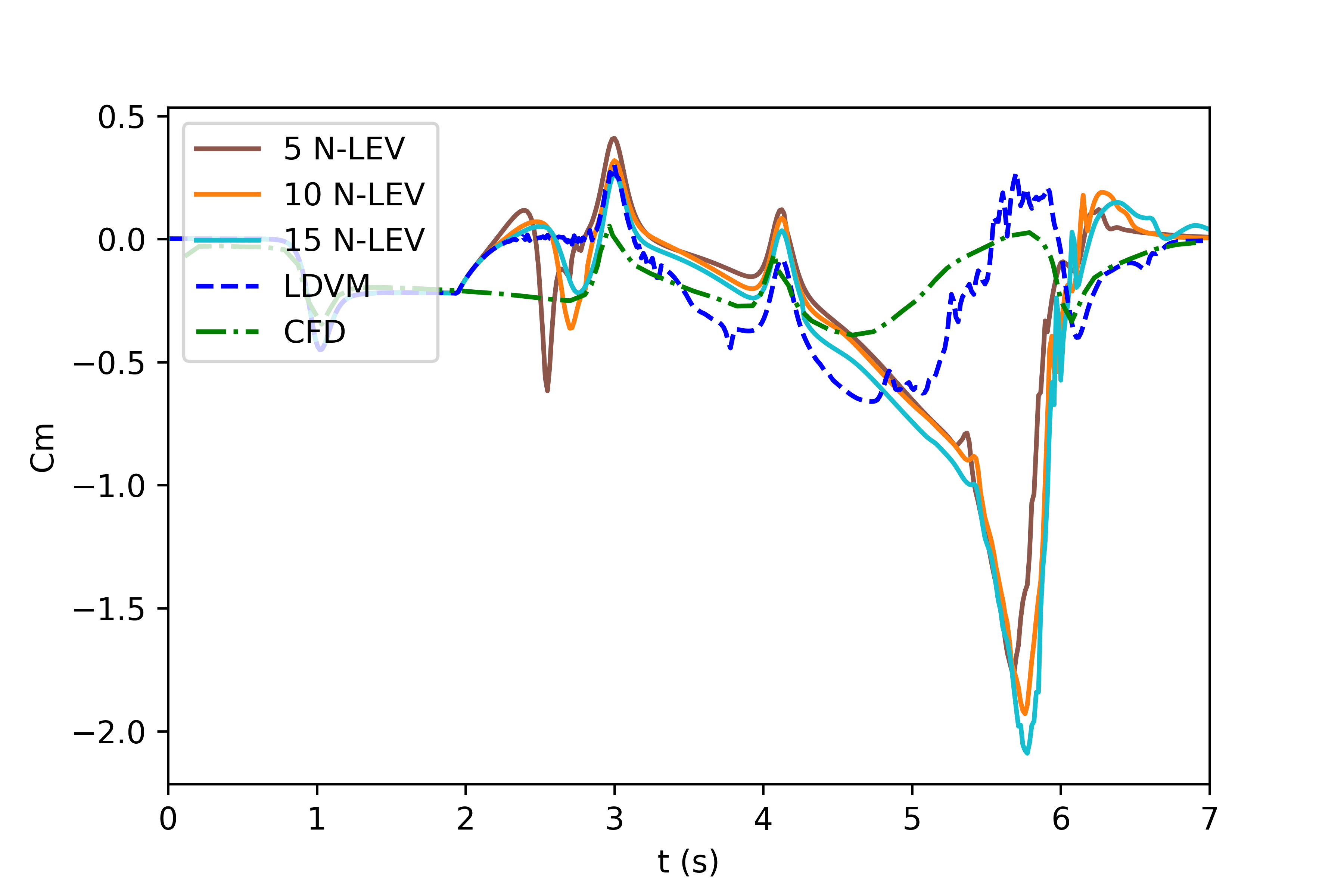}
\includegraphics[width=.49\textwidth]{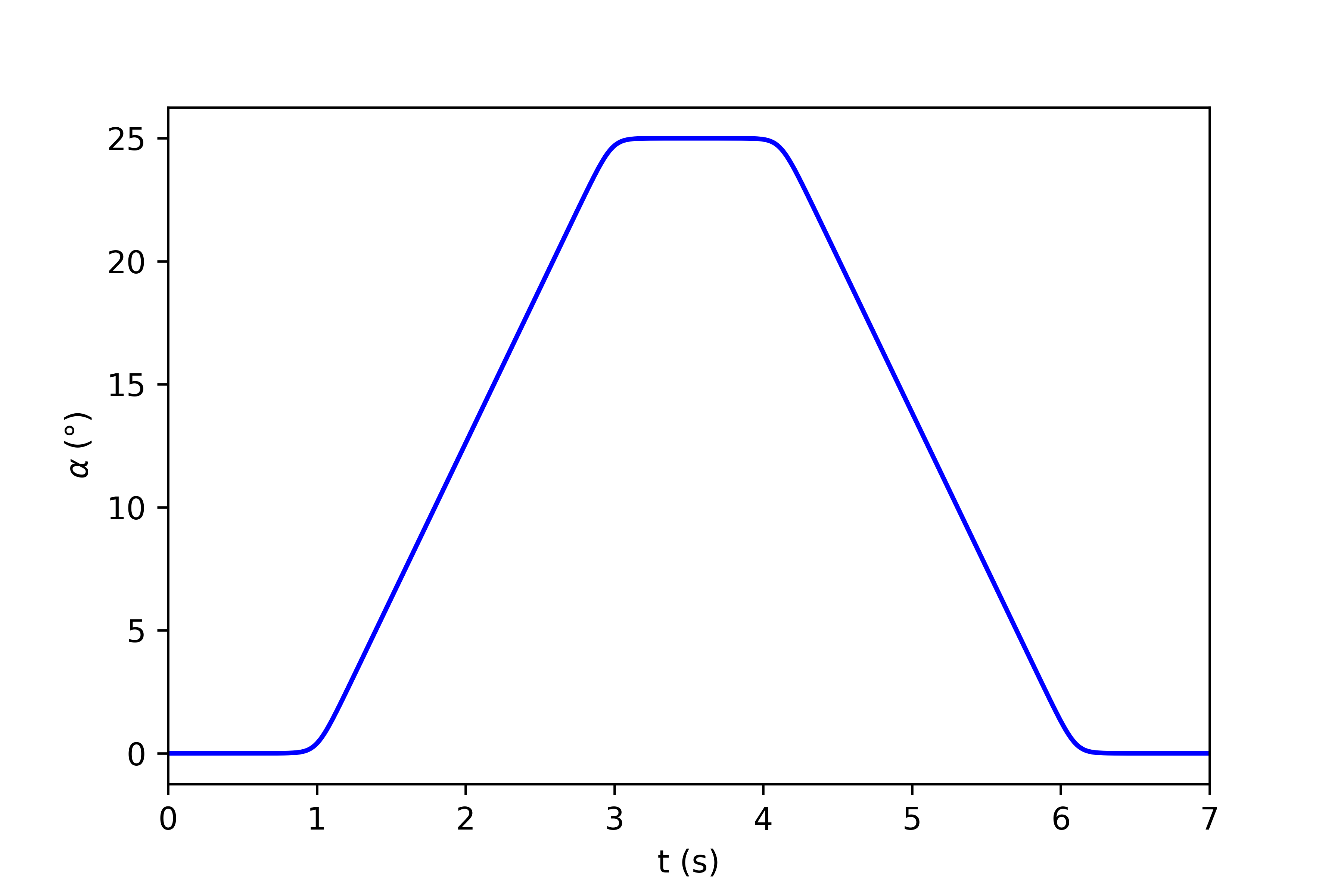}
  \caption{Case B: Comparison of $C_l$, $C_d$, $C_m$ and pitch angle from N-LEV LDVM (with 5, 10, 15 elements), classical LDVM and CFD.}
    \label{fig:resultskinem3}
\end{figure}

In this case we can see that very promising results occur before $t=4.5$, with close resemblance between the N-LEV method, LDVM and CFD results from Ramesh et al.\cite{ramesh2014discrete}. It is interesting to point that due to the constrained number of vortex elements the N-LEV method shows better numerical stability, as less vortex elements have to be discretised and accounted for. Again we see the 15-LEV method closely approaches the LDVM results until vortex detachment begins to occur. 

There is also an excessive computation of lift by N-LEV model in the early stages of LEV formation, even with similar vortex elements shapes, strengths and position. This may arise from a non-physical impulse produced by the merging of the vorticity produced at the L.E. with the vortex core of the 15-element N-LEV method, as shown by Widmann and Tropea\cite{widmann2015parameters} and Darakananda and Tropea\cite{darakananda2019versatile} and accounted for in their impulse-matching technique error correction. This is a key challenge in amalgamating vortex elements, and has to be accounted for as the method is developed to integrate detachment.  

\begin{figure}[h!]
\centering
\includegraphics[width=.33\textwidth]{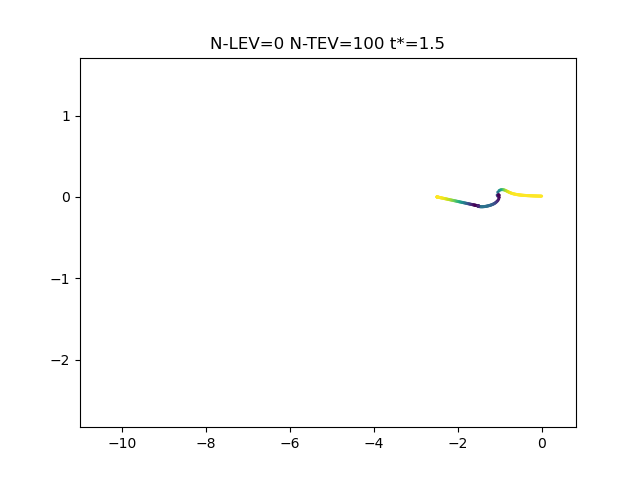}
\includegraphics[width=.33\textwidth]{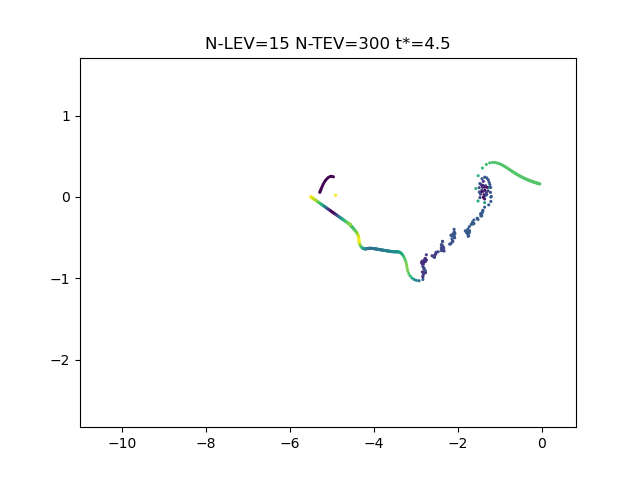}
\includegraphics[width=.33\textwidth]{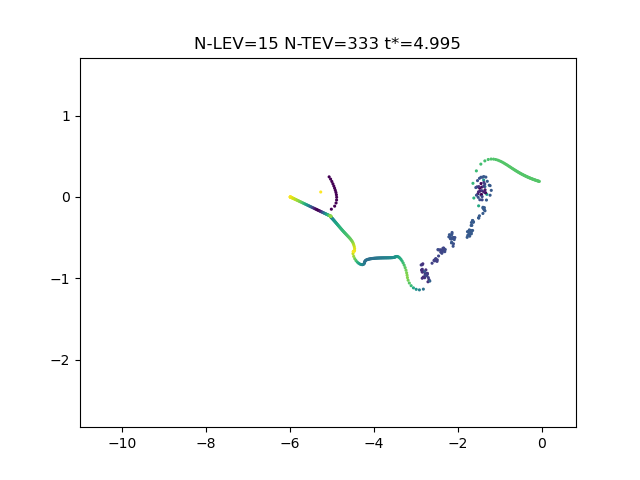}
\includegraphics[width=.33\textwidth]{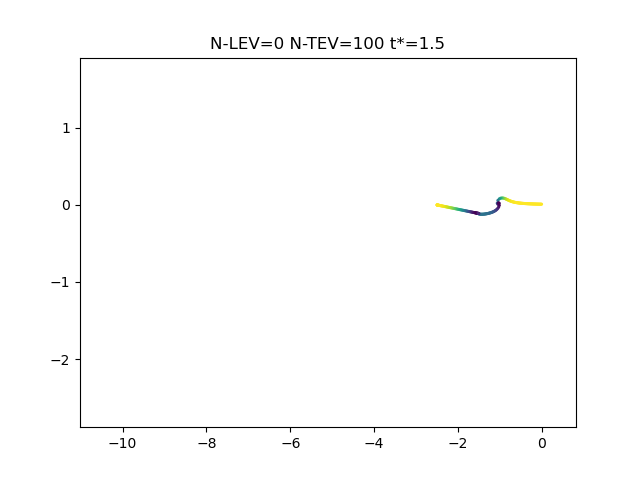}
\includegraphics[width=.33\textwidth]{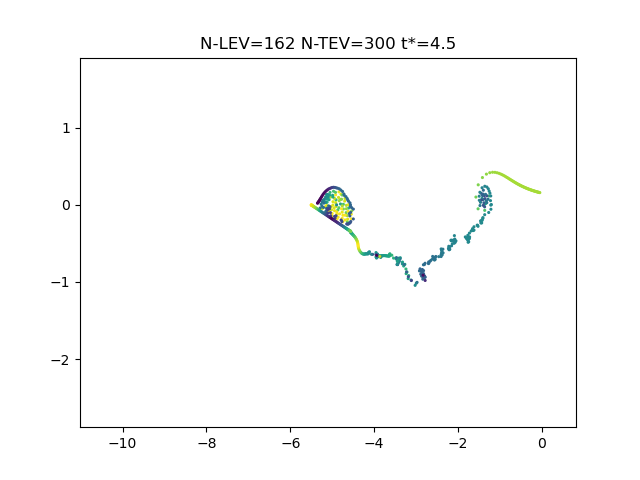}
\includegraphics[width=.33\textwidth]{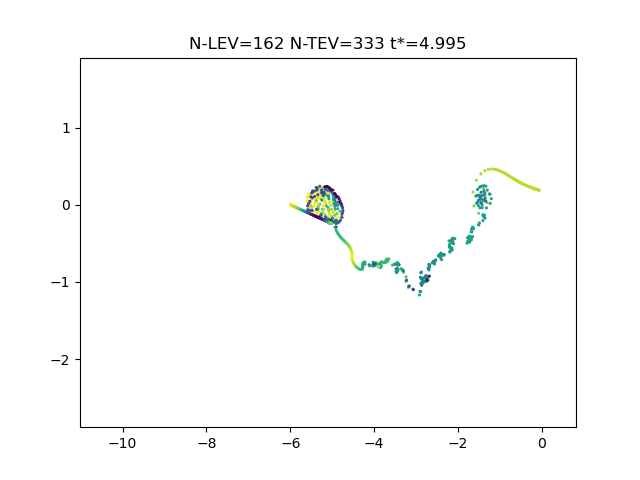}
  \caption{Case B: Vortex plots for N-LEV LDVM (with 15 elements) vs classical LDVM. The 15 LEV vortex element sheet is colored in magenta and the vortex core in yellow.}
    \label{fig:vortexplotskinem3}
\end{figure}
\FloatBarrier

\subsection{Demonstration of N-LEV LDVM for Case C}

Finally, an aggressive sinusoidal kinematic is presented with strong re-circulation and separation behind the airfoil. The same N-LEV limiting procedure is carried out in Section  \ref{sec:pitchramp45}, pure pitch-ramp-return kinematic, with no plunge component. The maximum amplitude is $\alpha_{max} = 90\degree$, and the pivot location the trailing edge. The kinematic case has a reduced frequency $k=0.4$, and results will be compared to CFD simulations by Ramesh et al.\cite{ramesh2014discrete}. In this case the LESP criterion is $LESP=0.14$, and the Eldredge ramp function used previously is tailored with a smoothing parameter $a=2$.

As the kinematic has significant flow separation, reducing the number of vortex elements could significantly accelerate the simulation. In this case in a 2.4 GHz processor desktop machine, compute time was found to be an average of $t_{compute}/t_{sim}/\Delta t = 0.024$ seconds per time step for the 10 element N-LEV method, and $t_{compute}/t_{sim}/\Delta t = 0.032$ for the original LDVM method. That is a 33\% increase in compute speed, putting the model closer to real-time performance if TEV vortices were limited. 

In this case very promising results in $C_l$ were obtained for the N-LEV method with $N= 10$, with unstable solutions being produced by both  $N= 5$ and  $N= 15$ after $t*=0.8$, which however did not extend to $C_d$, which was likely less affected due to the significance of vortex elements shed from the T.E. for the drag component. Moreover, results for $C_d$ were poorer than for $N= 10$ than both for $N= 5$ and $N= 15$ between $0.3<t*<0.55$.

In this aggressive kinematic significant flow separation occurs, with recirculation around the leading edge and trailing edge rapidly occurring after LEV and TEV onset. Hence the component of the airfoil forces produced by the TEV are significant. 

Future efforts to limit the N-LEV models number of TEV vortices will have to take this into account. This also points out to the importance of selecting a tailored N parameter adapted to the vortex sheet and coherent structures of interest, analogous to the importance of the parameters $B_f$ and $T_{min}$ proposed by Darakananda and Eldredge\cite{darakananda2019versatile}. As the performance of the N-LEV model is non-obvious and may vary across airfoil forces, it will be of interest to study the quantitative optimization of the $N$ parameter, and the correlation between accuracy and the input kinematics to the model. 

\begin{figure}[h!]
\centering
\includegraphics[width=.49\textwidth]{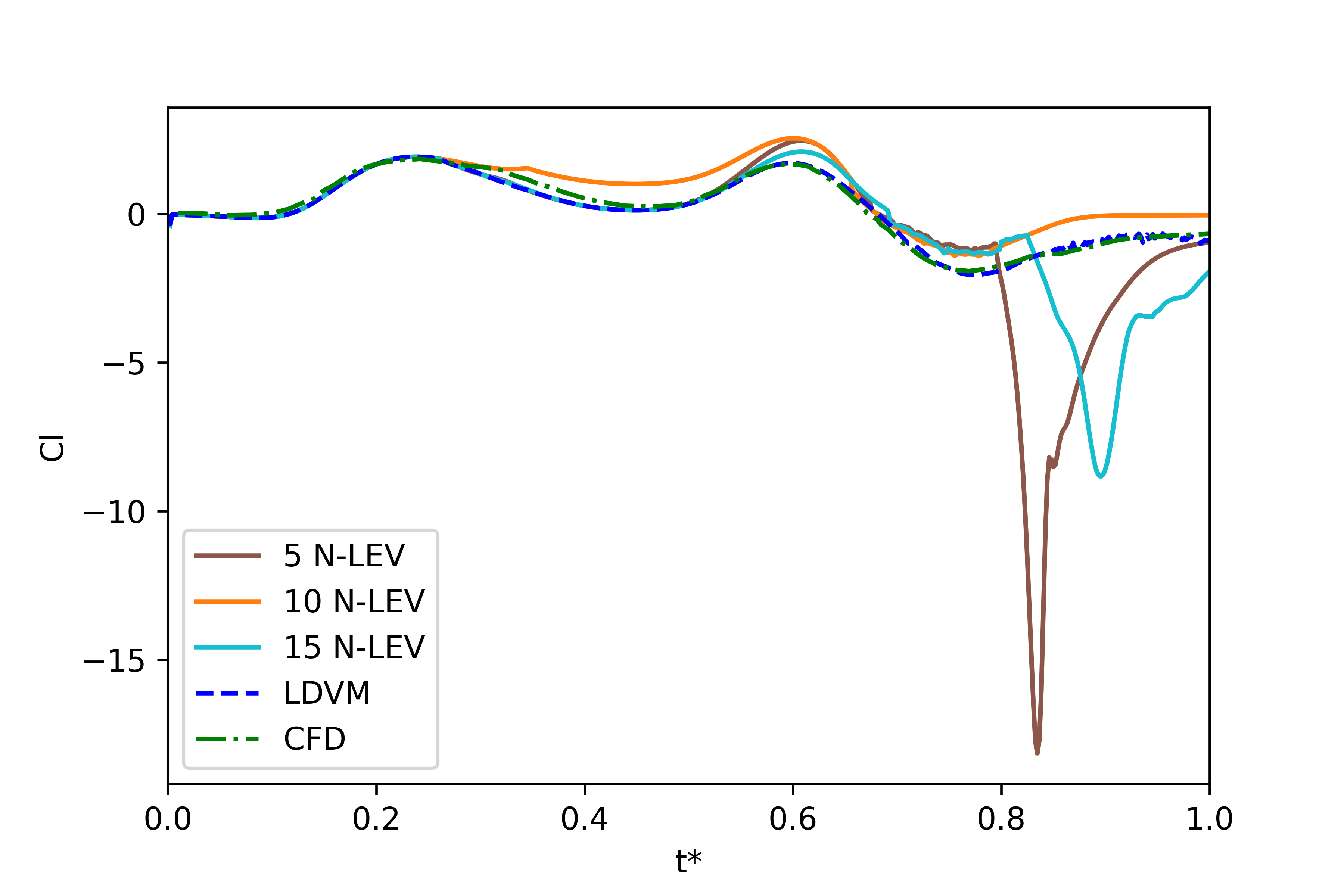}
\includegraphics[width=.49\textwidth]{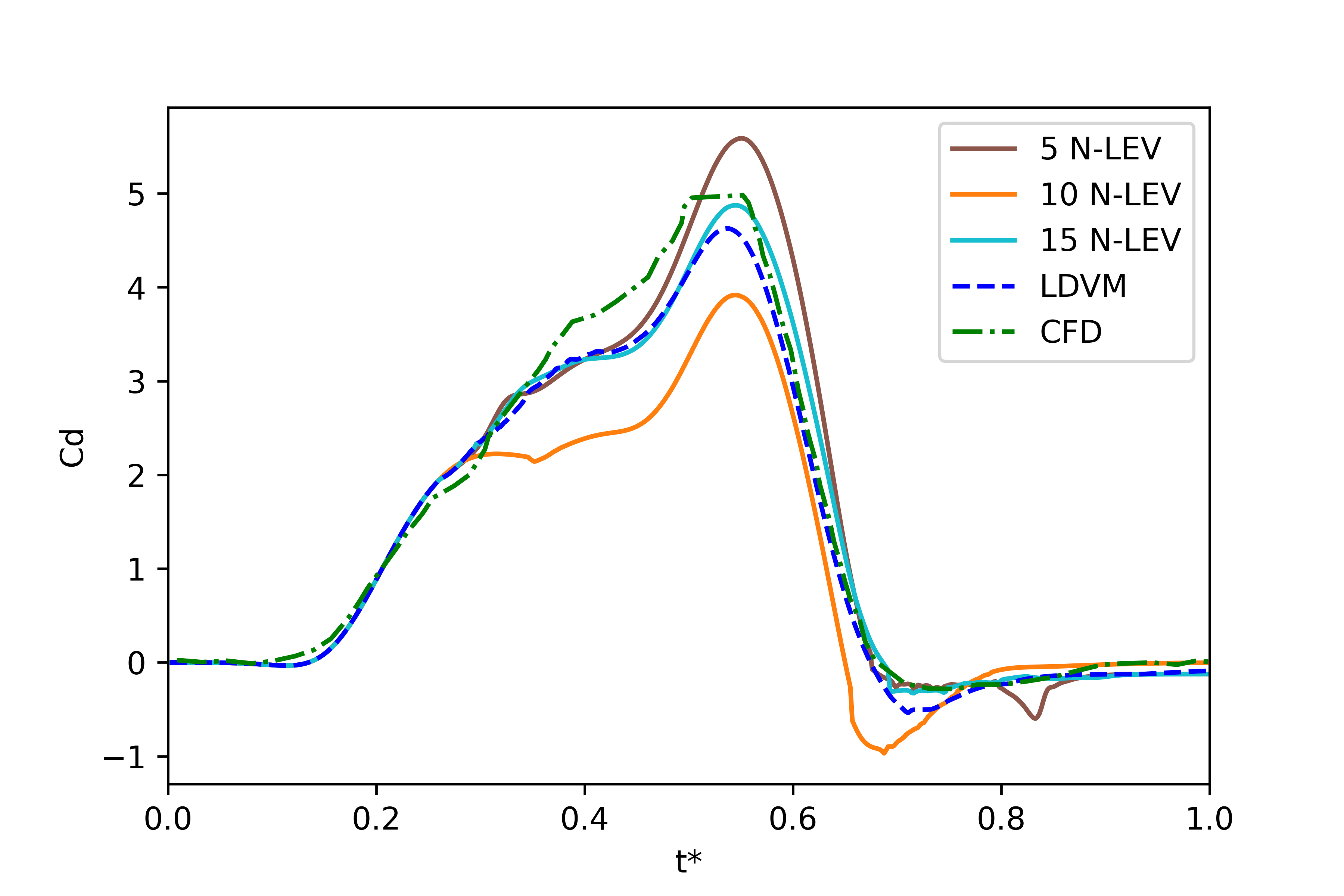}
\includegraphics[width=.49\textwidth]{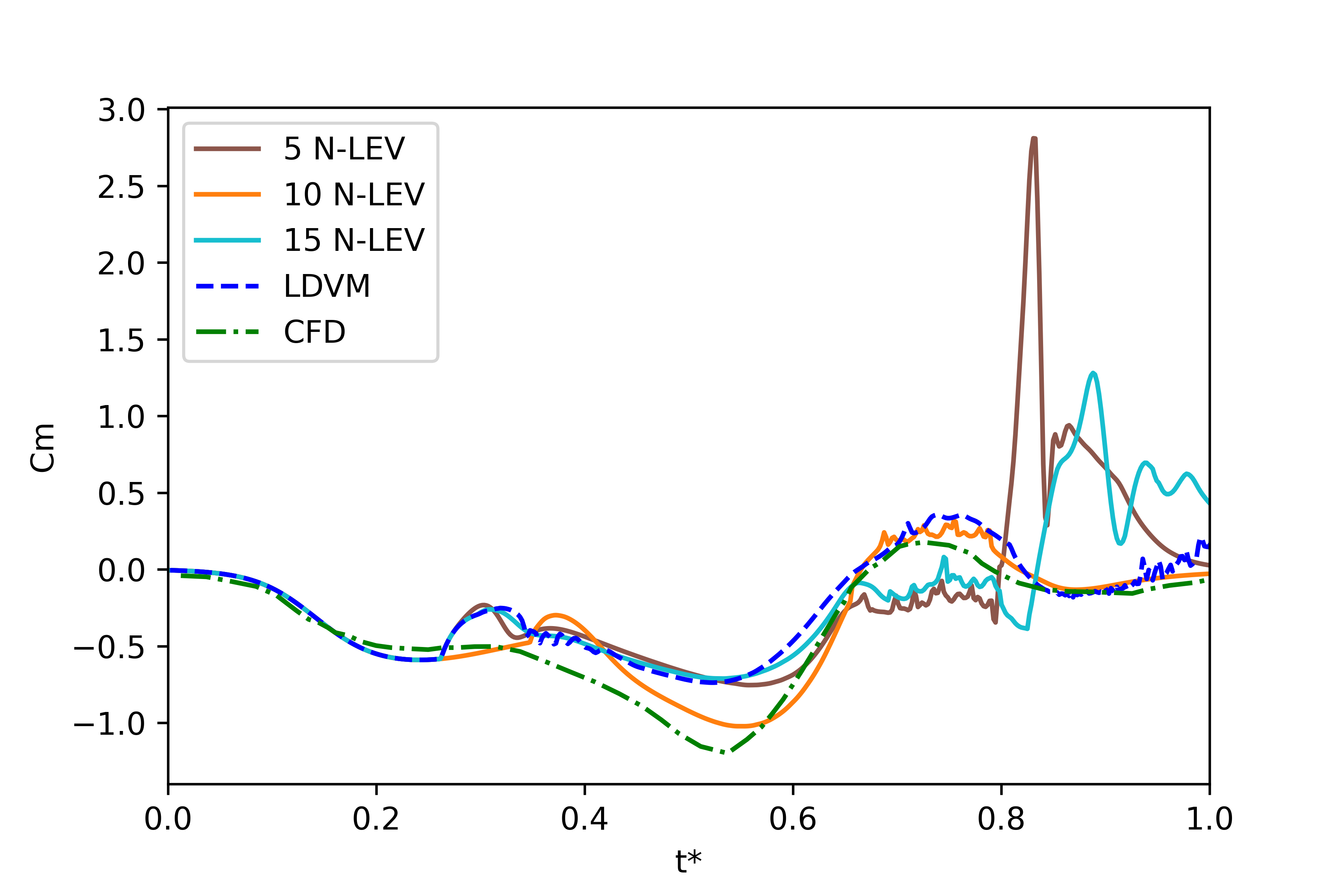}
\includegraphics[width=.49\textwidth]{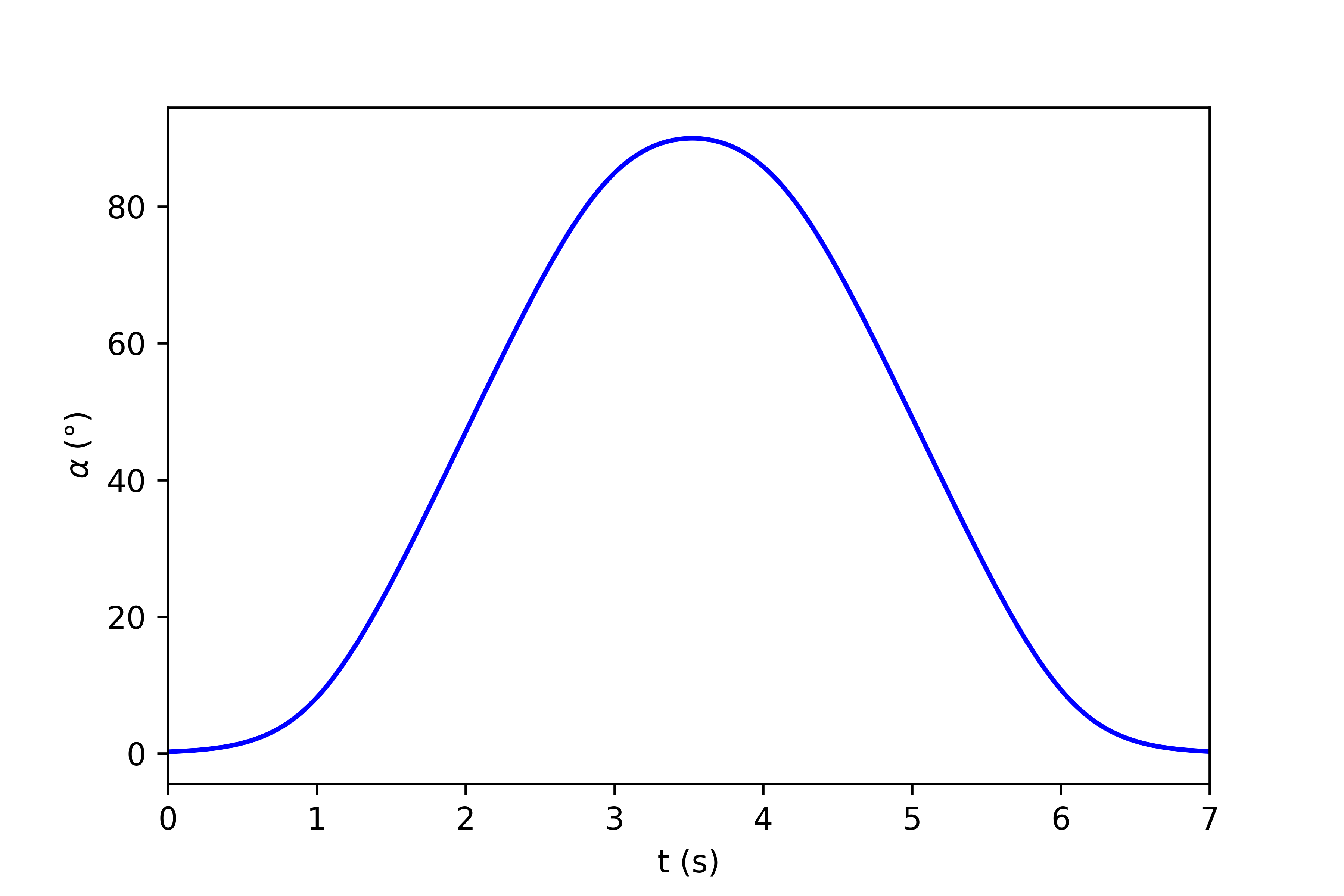}
  \caption{Case C: Comparison of $C_l$, $C_d$, $C_m$ and pitch angle from N-LEV LDVM (with 5, 10, 15 elements), classical LDVM and CFD.}
    \label{fig:resultskinem4}
\end{figure}

\begin{figure}[h!]
\centering
\includegraphics[width=.32\textwidth]{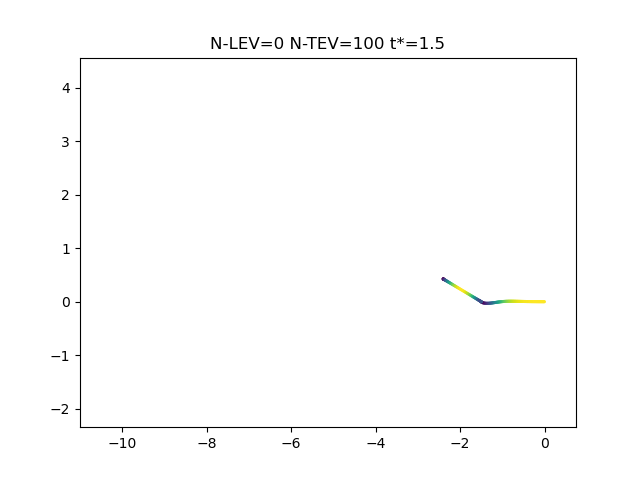}
\includegraphics[width=.32\textwidth]{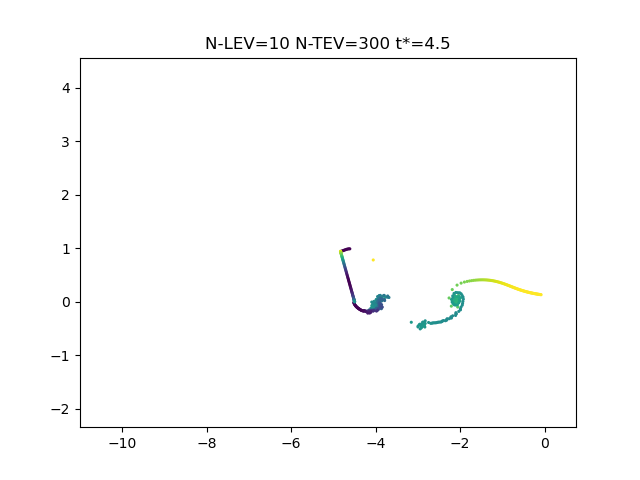}
\includegraphics[width=.32\textwidth]{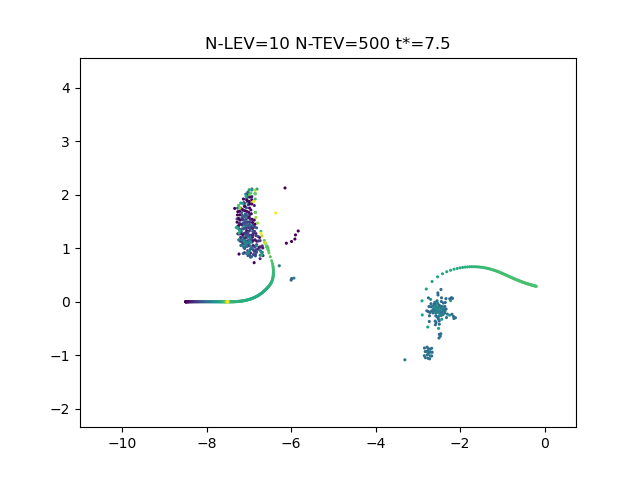}
\includegraphics[width=.32\textwidth]{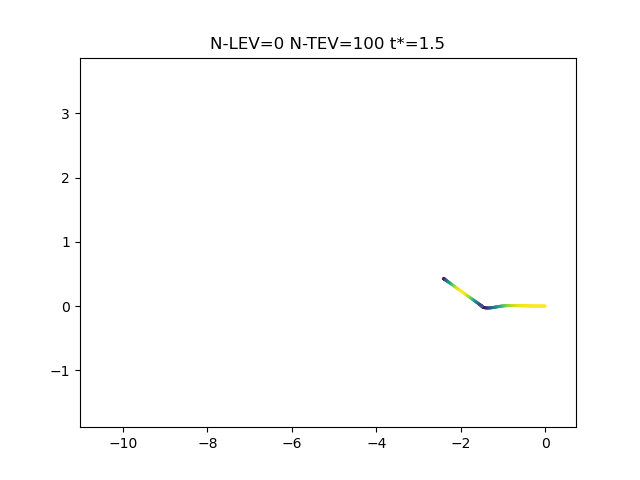}
\includegraphics[width=.32\textwidth]{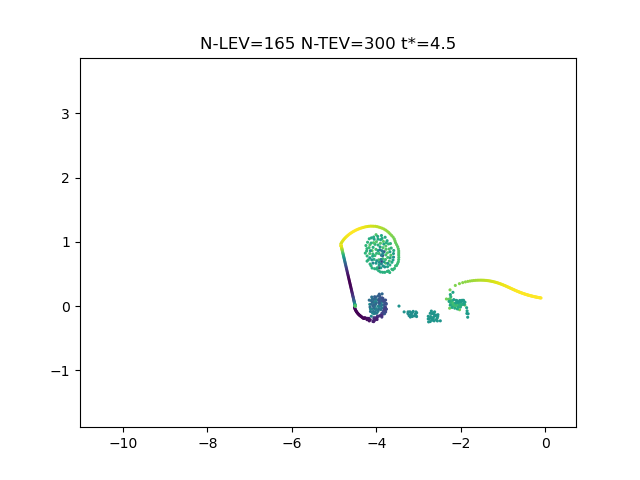}
\includegraphics[width=.32\textwidth]{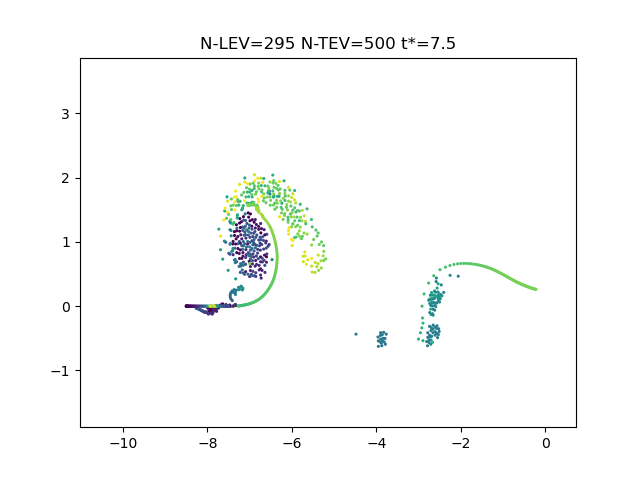}

  \caption{Case C: Vortex plots for N-LEV LDVM (with 15 elements) vs classical LDVM. The 15 LEV vortex element sheet is colored in magenta and the vortex core in yellow.}
    \label{fig:vortexplotskinem4}
\end{figure}

Beyond $t*>=0.7$ the 10-LEV method over-predicts lift generation by the airfoil, the cause of this becomes evident when observing the vortex element plots in Fig. \ref{fig:vortexplotskinem4}. As the airfoil descends from $\alpha_{max}$ back to $\alpha=0$, some LEV elements generated in the final steps of the kinematic are convected below the airfoil surface rather than into the wake. This causes a low-pressure zone that generates a negative lift component, which is visibly absent from the N-LEV method both in the vortex plots in Fig. \ref{fig:vortexplotskinem4} and in the $C_l$ plot in Fig. \ref{fig:resultskinem4}. 

\FloatBarrier

\subsection{Leading Edge Vortex Detachment Criterion 1: Trailing Edge Flow Reversal}

Results in the previous section showed that the N-LEV model performed well only until detachment of the LEV from the airfoil surface. Hence the identification of detachment is critical in accurately modeling the flow beyond this time. In this section, the ability of N-LEV LDVM to identify detachment from the location of the rear stagnation point (shown as a green diamond in Fig. \ref{fig:flowtopology}) is studied. We aim to understand the effect of limiting the number of vortex elements on the location of the stagnation point on the airfoil, and more generally the effect on the overall flow topology. 

\begin{figure}[h!]
\centering
\includegraphics[width=.65\textwidth]{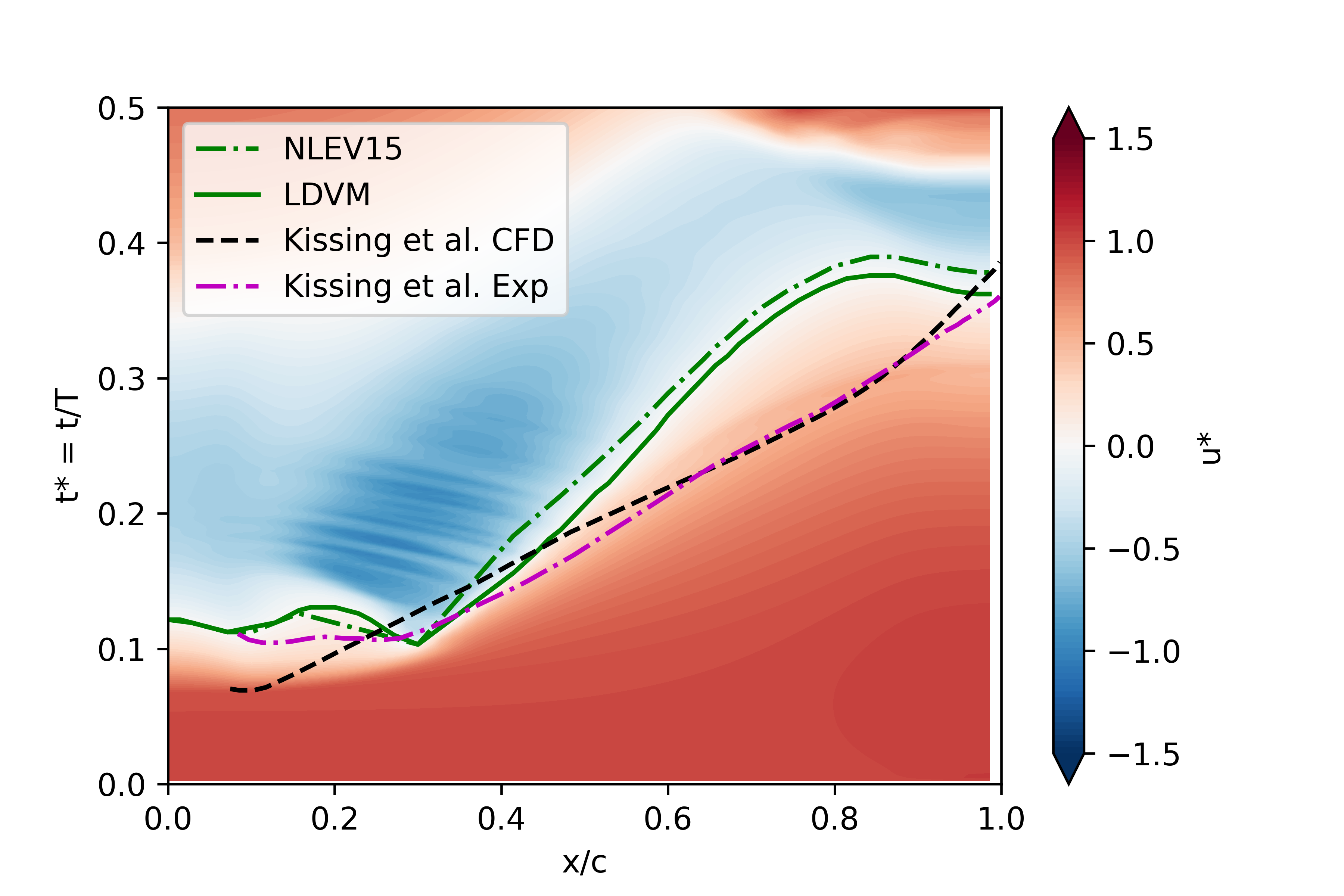}
  \caption{Comparison of stagnation point location (x-axis) vs nondimensional time (y-axis) between N-LEV LDVM, classical LDVM and results from Kissing et al.\cite{kissing2020insights}. Non-dimensional time (t*) against position (u* = $\frac{u}{U_{inf}}$) contours from -1.5 (blue) to 1.5 (red) are shown from the LDVM method.}
    \label{fig:surfacevelsldvm}
\end{figure}

Results from Widmann \& Tropea\cite{widmann2015parameters}, and recently Kissing et al. \cite{kissing2020insights} point to several detachment mechanisms for the LEV. This method will study their flow-reversal at the trailing edge as a criterion for detachment. We have replicated the kinematics of Kissing et al. \cite{kissing2020insights} of a pitching and plunging airfoil with a plunge amplitude of $h=0.65c$, pitch amplitude of $12.67\degree$, reduced frequency $k=\pi c/U_{\infty}T = 0.48$, $St=hc/U_{\infty}T = 0.1$, where $T$ is the period and $U_{\infty}$ the free stream velocity. As seen in Fig. \ref{fig:surfacevelsldvm}, the velocity tangential to the airfoil surface can be determined from experimental and numerical data. The green line shows the evolution of the rear stagnation point, which causes flow reversal when it reaches the trailing edge, that is $x/c = 1$. 

Of central importance to compute a detachment criteria based on Kissing et al. \cite{kissing2020insights} results is the calculation of the surface velocities in the airfoil. These can be obtained from the LDVM method by Eq. \ref{equation:surfacevelsu} for the upper surface and Eq. \ref{equation:surfacevelsl} for the lower surface. Depending on the kinematics and where the LEV is generated one or the other will be required to calculate the location of the stagnation point on the airfoil.

\begin{equation}
    V_{tu} = U cos(\alpha) + \dot{h}sin(\alpha) + \left(\frac{\delta\phi_{lev}}{\delta x}\right)_{u} + \left(\frac{\delta\phi_{tev}}{\delta x}\right)_u + \left(\frac{\delta\phi_{b}}{\delta x}\right)_u
\label{equation:surfacevelsu}
\end{equation}

\begin{equation}
    V_{tu} = U cos(\alpha) + \dot{h}sin(\alpha) + \left(\frac{\delta\phi_{lev}}{\delta x}\right)_l + \left(\frac{\delta\phi_{tev}}{\delta x}\right)_l + \left(\frac{\delta\phi_{b}}{\delta x}\right)_l
\label{equation:surfacevelsl}
\end{equation}


The authors of this paper have managed to obtain results for the evolution of this rear stagnation point from the LDVM method by extracting the velocity tangential to the airfoil on its surface\cite{rival2014characteristic}, further demonstrating that this phenomena is not strictly viscous, and that it can be used to augment low-order methods. 

However, as can be seen in Fig. \ref{fig:surfacevelsldvm} the tracing of this stagnation point through the kinematic by the inviscid LDVM method is poor in $0.1<t*<0.3$, with significant deviations from Kissing et al.\cite{kissing2020insights} numerical and experimental results. As seen Fig. \ref{fig:surfacevelsldvm}, the inviscid models predict a faster approach of the stagnation point in the surface of the airfoil towards the trailing edge than numerical and experimental results do. 

Physical mechanisms absent in the LDVM and N-LEV models could be behind this. Viscous phenomena occurring in the boundary layer not captured by the inviscid model may delay the progress of this stagnation point through the surface of the airfoil by causing steeper velocity gradients from the surface that inhibit the growth of the LEV. 

However, in the context of the development of an augmented LDVM method, the time instant of interest is that when trailing edge flow reversal occurs, that is $u=0$ at $x/c=1$. For this, the LDVM and N-LEV methods show close agreement with Kissing et al.\cite{kissing2020insights} results, showing promise in deriving a detachment criterion. This point is predicted to occur by LDVM at $t*=0.362$ and by the N-LEV model at $t*=0.378$, whilst experimental results of Kissing et al.\cite{kissing2020insights} predict this point at $0.38<t*<0.39$ depending on the setup used. Hence the vortex-element limited N-LEV LDVM model shows closer agreement that the LDVM model at this point. The fact that this is the case points out that trailing edge flow reversal is a purely inviscid LEV detachment mechanism.  


The arrival of this rear stagnation point to the trailing edge can be used in the N-LEV LDVM method to halt the accumulation of vorticity in the LEV, causing it to detach and be shed downstream into the wake, capturing this large vortex feature in the low-order method. The next N-LEV would be initiated at this stage. This model will be pursued in future research.

The ability of a fast method to calculate the evolution of this stagnation point (LDVM simulations can run several convective times in less than 1s in a desktop computer), allows for the possibility of testing different kinematics and airfoil shapes that halt the arrival of this point to the trailing edge, maximizing the lift augmentation effect from the LEV, as insect fliers do naturally \cite{wootton1999flies}.

\subsection{Leading Edge Vortex Detachment Criterion 2: Maximum circulation of the LEV}

The second detachment mechanism of interest proposed by Widmann and Tropea\cite{widmann2015parameters}, as well as  experimentally and numerically studied by Kissing et al. \cite{kissing2020insights}, is a maximum LEV circulation $\Gamma_{LEV}$ which can be converted into a vortex Reynolds number. 

\begin{figure}[h!]
    \centering
    \includegraphics[width=0.65\textwidth]{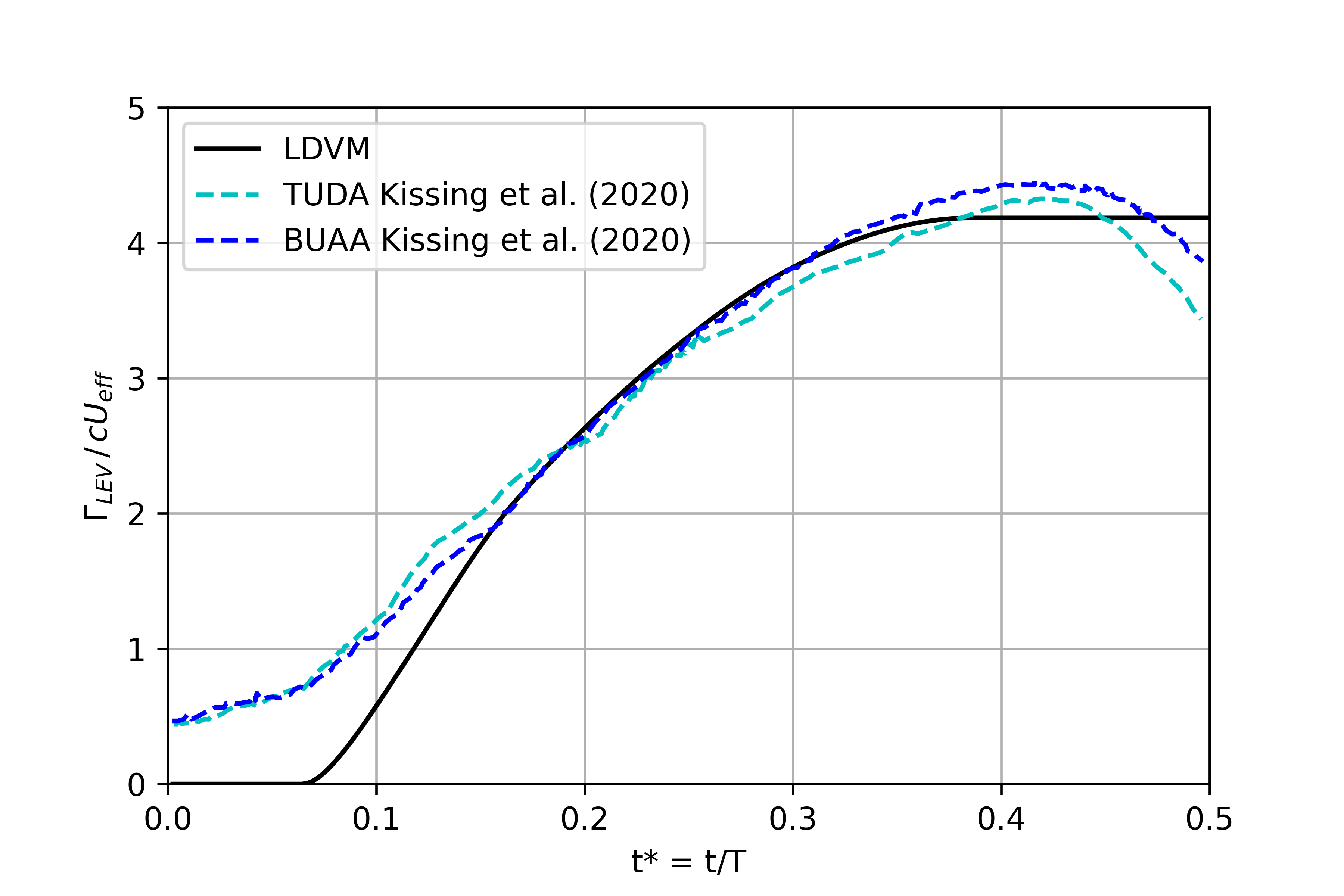}
    \caption{Comparison of accumulated leading edge vortex circulation  vs non-dimensional time between Kissing et al. (2020) and LDVM, $U_{eff}=U_{\infty}$}
    \label{fig:maxcirculationkissing}
\end{figure}

The same kinematics from Kissing et al. \cite{kissing2020insights} used in the previous section are used here. Our results in Fig. \ref{fig:maxcirculationkissing} closely resemble those obtained in their experimental context, pointing out to this maximum LEV circulation that limits vortex growth and causes further detachment. 

In fact, this seems to be captured by the LESP criterion in the LDVM method proposed by Ramesh et al.\cite{ramesh2014discrete} which halts the shedding of vortex elements from the leading edge when below the required threshold, which is evident in Fig.\ref{fig:maxcirculationkissing} as the maximum constant circulation reached by the LEV $\Gamma_{LEV}/cU_{eff}=4.2$. Hence this vorticity accumulation mechanism may be related to the capacity of the leading edge to sustain a certain degree of suction. Recent developments in the modeling of this criterion as a function of time by Martinez Carmena et al. \cite{martinez2022modulation} will be implemented in future work. This mechanism also shows promise for the N-LEV method as leading edge vortex generation is independent on the number of vortex elements chosen to cap N-LEV method. 

\FloatBarrier

\section{Conclusions}
The paper aimed to study the effects of constraining the number of vortex elements shed from the leading edge in the LDVM. Whilst performance is encouraging and physical features are recreated in leading edge vortex formation, excessive and nonphysical accumulation of circulation by the LEV core limits the models' accuracy beyond LEV detachment. 

Two physical shedding mechanisms from literature were analyzed as proposals to enhance the N-LEV method by reintroducing physical detachment absent from the trivial vortex element limiting. The independence of these mechanisms was studied, as well as the effect of limiting the number of vortex elements on them.  


Initial results presented are promising with cases in which the N-LEV LDVM method agrees closely qualitatively with $C_l$, $C_d$ and $C_m$, also having up to a 28\% reduction in computational expense. Error diverges rapidly beyond LEV detachment, requiring the introduction of the physical detachment mechanisms. In certain kinematics with high degrees of flow separation, the N-LEV LDVM method strongly approximates the original LDVM method and CFD data. 

Future work on the method will involve introducing the mechanisms to trigger leading edge vortex detachment by halting the accumulation of vorticity and proposing similar mechanisms for the trailing edge vortex.

\section*{Acknowledgments}
The authors gratefully acknowledge the support of the College of Science and Engineering, University of Glasgow through a DTA scholarship. 


\bibliography{sample}

\end{document}